\documentclass[a4paper,11pt]{article}
\pdfoutput=1

\usepackage{jcappub}

\newcommand{\by}{{\boldsymbol{y}}}
\newcommand{\bC}{{\boldsymbol{C}}}

\newcommand{\bW}{{\boldsymbol{W}}}

\newcommand{\bdelta}{{\boldsymbol{\delta}}}
\newcommand{\bDelta}{{\boldsymbol{\Delta}}}

\newcommand{\bLambda}{{\boldsymbol{\Lambda}}}
\newcommand{\btheta}{{\boldsymbol{\theta}}}
\newcommand{\bTheta}{{\boldsymbol{\Theta}}}
\newcommand{\bmu}{{\boldsymbol{\mu}}}
\newcommand{\brho}{{\boldsymbol{\rho}}}
\newcommand{\bmM}{{\boldsymbol{\mathcal{M}}}}
\newcommand{\mL}{{\mathcal{L}}}

\newcommand{\ML}{{\text{ML}}}
\newcommand{\chisquaredred}{{\chi_{\text{red}}^2}}
\newcommand{\psisquaredred}{{\psi_{\text{red}}^2}}
\newcommand{\E}{{\text{E}}}
\newcommand{\Var}{{\text{Var}}}
\newcommand{\Cov}{{\text{Cov}}}
\newcommand{\Corr}{{\text{Corr}}}

\usepackage{mwe}
\usepackage{multirow}

\newlength{\tempdima}
\newcommand{\rowname}[1]
{\rotatebox{90}{\makebox[\tempdima][c]{\textbf{#1}}}}

\usepackage[T1]{fontenc} 

\title{A new goodness-of-fit statistic and its application to 21-cm cosmology}

\author[a,b,1]{K. Tauscher\note{Corresponding author.}}
\author[a,c]{, D. Rapetti}
\author[a]{, and J.O. Burns}

\affiliation[a]{Center for Astrophysics and Space Astronomy,\\University of Colorado Boulder,\\
389 UCB, Boulder, CO 80309, USA}
\affiliation[b]{Department of Physics,\\University of Colorado Boulder,\\390 UCB, Boulder, CO 80309, USA}
\affiliation[c]{NASA Ames Research Center,\\Moffett Blvd, Mountain View, CA 94035, USA}

\emailAdd{Keith.Tauscher@colorado.edu}
\emailAdd{David.Rapetti@colorado.edu}
\emailAdd{Jack.Burns@colorado.edu}

\abstract{The reduced chi-squared statistic is a commonly used goodness-of-fit measure, but it cannot easily detect features near the noise level, even when a large amount of data is available. In this paper, we introduce a new goodness-of-fit measure that we name the reduced psi-squared statistic. It probes the two-point correlations in the residuals of a fit, whereas chi-squared accounts for only the absolute values of each residual point, not considering the relationship between these points. The new statistic maintains sensitivity to individual outliers, but is superior to chi-squared in detecting wide, low level features in the presence of a large number of noisy data points. After presenting this new statistic, we show an instance of its use in the context of analyzing radio spectroscopic data for 21-cm cosmology experiments. We perform fits to simulated data with four components: foreground emission, the global 21-cm signal, an idealized instrument systematic, and noise. This example is particularly timely given the ongoing efforts to confirm the first observational result for this signal, where this work found its original motivation. In addition, we release a Python script dubbed \texttt{psipy} which allows for quick, efficient calculation of the reduced psi-squared statistic on arbitrary data arrays, to be applied in any field of study.}

\begin{document}
\maketitle
\flushbottom

\section{Introduction}

Observing the highly-redshifted, sky-averaged 21-cm spectrum of the hyperfine transition of neutral hydrogen's ground state, usually referred to as the global 21-cm signal, is currently a key objective in radio astronomy and for the astrophysics community at large. It is information-rich, having the capability to constrain the properties of the first stars and X-ray emitting black holes through the thermal and reionization histories of the early Universe ($\sim10-200$ MHz) \cite{cohen-2017,mirocha-2018}. At very low frequencies ($\sim10-30$ MHz), it can even be used to probe for signs of exotic physics at the end of the Dark Ages, free of the astrophysics of the first luminous objects, such as unexplained cooling beyond the maximum adiabatic level set by cosmic expansion. This is specially motivated by the recent results from the low-band instrument of the Experiment to Detect the Global Epoch of Reionization (EoR) Signature (EDGES) \citep{monsalve-2017,monsalve-2018}. They report an absorption feature at 78 MHz \cite{bowman-2018} which, if taken to correspond to the Cosmic Dawn trough, is $2-3$ times larger in amplitude than was predicted possible without additional cooling or a stronger radio background. This is presently an intensely investigated topic in the field \cite{barkana-2018, ewall-wice-2018, fraser-2018, kovetz-2018}. Some have expressed concerns about the modeling of the foregrounds present alongside the trough \cite{hills-2018}, leading to questions that the new statistic presented in this paper may help resolve.

 Other experiments attempting to measure this signal include the Shaped Antenna measurement of the background RAdio Spectrum (SARAS) \citep{singh-2018}, the Sonda Cosmol\'ogica de las Islas para la Detecci\'on de Hidr\'ogeno Neutro (SCI-HI) \citep{voytek-2014}, the Large-aperture Experiment to detect the Dark Ages (LEDA) \citep{price-2018}, Probing Radio Intensity at high-z from Marion (PRIzM) \citep{philip-2018}, and the Cosmic Twilight Polarimeter (CTP) \citep{nhan-2018b}.~In \cite{burns-2017}, the authors suggested a space-based observational strategy to measuring the 21-cm signal, which is currently being employed in the NASA-funded concept study for the Dark Ages Polarimeter PathfindER (DAPPER), a lunar-orbiting SmallSat experiment to be deployed from the upcoming Lunar Gateway. All global 21-cm signal experiments see a combination of four data components: beam-weighted foregrounds, the desired global 21-cm signal, instrumental systematics, and noise. Generally, the residuals resulting from fitting these data contain wide-band structure because the foreground, systematic, and signal --- as well as the functions used to model them --- are in most cases spectrally smooth and delocalized. The challenge addressed in this paper is that these wide-band, low-level residual features normally complicate the essential task of measuring the goodness-of-fit of chosen models.

A traditional quantity used to measure goodness-of-fit is the reduced chi-squared statistic $\chisquaredred$, which is proportional to the mean-square, error-weighted difference between the data and the model evaluated at the fit parameters \citep{pearson-1900,cochran-1952}. It is useful because it consistently uses all data points and its distribution is well known when the data are sufficiently modeled compared to the (known) noise level. Because each residual value is individually squared, $\chisquaredred$ is excellent for uncovering the presence of features whose amplitudes far exceed the noise level, even if they exist in only a small segment of the residual. On the other hand, others have noted failings of $\chisquaredred$ and suggested additional methods, such as tests involving quadratic forms of order statistics \cite{shapiro-1965} or empirical distribution functions \cite{stephens-1974} (see also, e.g., \cite{d'agostino-1986,genest-2009,gonzalez-manteiga-2013} for reviews on goodness-of-fit tests). Nonetheless, $\chisquaredred$ continues to be very commonly employed and is thus a useful reference for comparisons in this paper.

When wide-band features in residuals of fits to 21-cm cosmology data are large in amplitude (compared to the noise level), they are easily identified by $\chisquaredred$. However, if the amplitude of the feature is comparable to the noise level as happens often in unbinned data or when the model of the data is nearly sufficient, then the sensitivity of $\chisquaredred$ drops significantly because it only measures the magnitude of the residual points themselves, ignoring any relation between the different points; it can indeed be thought of as the $0^{\text{th}}$ order correlation of the residual. In this paper, we define the reduced psi-squared statistic $\psisquaredred$, which reflects the amplitude of all nonzero correlation orders. In general, this new statistic is vastly superior to $\chisquaredred$ in measuring whether large-scale low-amplitude features exist in a given data vector. This makes it especially useful in analyzing 21-cm cosmology data.

In Section~\ref{sec:methods}, we give background about the standard $\chisquaredred$ statistic, define the new $\psisquaredred$ statistic, and describe their joint and marginalized distributions in various cases. In Section~\ref{sec:21-cm-cosmology}, we present an idealized example of the application of $\psisquaredred$ to fitting simulated data from 21-cm experiments. We conclude in Section~\ref{sec:discussion} by discussing the promise of and prospects for future work on the new $\psisquaredred$ statistic, as well as its further usefulness in fields like 21-cm cosmology.

\section{Methods} \label{sec:methods}

In this section, we lay out our methods for fitting (Section~\ref{sec:fitting}) and describe the standard goodness of fit statistic $\chisquaredred$ (Section~\ref{sec:chi-squared}) before defining our new statistic $\psisquaredred$ (Section~\ref{sec:definitions}) and exploring its distribution in the presence of pure white noise (Section~\ref{sec:null-hypothesis}) and its sensitivity to non-random components in residuals (Sections~\ref{sec:nonrandom}~and~\ref{sec:random-and-nonrandom}). In Section~\ref{sec:hypothesis-test}, we propose a formal hypothesis test with which to determine whether or not the value of $\psisquaredred$ indicates that residuals are not purely noise.

  \subsection{Fitting procedures} \label{sec:fitting}
  
  Consider the problem of fitting a data vector (e.g. a spectrum or group of sky-averaged spectra) $\by$, in the presence of Gaussian noise with covariance $\bC$, and a model $\bmM(\btheta)$ with parameters $\btheta$ in a space $\bTheta$. In this situation, a likelihood function $\mL(\btheta)$ is usually defined as
  \begin{equation}
    \mL(\btheta) \propto \exp{\left\{-\frac{1}{2}\bdelta^T\bC^{-1}\bdelta\right\}} \ \ \text{ where } \ \ \bdelta=\by-\bmM(\btheta). \label{eq:likelihood}
  \end{equation}
  One may explore the posterior parameter probability density through numerical sampling with this likelihood function. But, for this to be a meaningful task, it must first be determined that the model being used is reasonable given prior knowledge of the system, which is mostly determined by the single best fit. For this reason, in this paper, we discuss only the maximum likelihood value of $\btheta$, not the noise-driven scatter around this value. To find this maximum likelihood parameter vector,
  \begin{equation}
    \btheta_{\ML} \equiv \underset{\btheta\in\bTheta}{\text{argmax}}\ \mL(\btheta),
  \end{equation}
  when performing fits (see Section~\ref{sec:21-cm-cosmology}), we use two different techniques included in the publicly available \texttt{pylinex} code,\footnote{\label{footnote:pylinex-link}\url{https://bitbucket.org/ktausch/pylinex}} both of which require that we define $\ln{\mL(\btheta)}$. The first technique, used if the model $\bmM(\btheta)$ is linear (i.e. it can be written as $\bmM(\btheta)=\bLambda\btheta$ for a matrix $\bLambda$), is that of analytical solution using linear algebra with the equations described in \cite{tauscher-2018}. For the second technique, which can be employed on any likelihood, we use a numerical minimization algorithm provided by \texttt{scipy.optimize}\footnote{\url{https://www.scipy.org}} to minimize $-\ln{\mL(\btheta)}$ by ascending the gradient $\boldsymbol{\nabla}_\btheta\ln{\mL(\btheta)}$. To be robust in the case of noisy likelihoods or bad initial guesses, we perform this process many times from different starting positions. This is similar to the ``basin hopping'' method of \cite{wales-1997}, which was created to explore rugged energy surfaces with a large number of local minima, differing in that the iterations of gradient descent here are performed independently.

\subsection{Goodness-of-fit and $\chisquaredred$ statistic} \label{sec:chi-squared}

  A standard statistical goodness-of-fit metric used for Gaussian likelihoods is the reduced chi-squared statistic, $\chisquaredred$, given by
  \begin{equation}
    \chisquaredred = \frac{\bdelta_{\ML}^T\bC^{-1}\bdelta_{\ML}}{N-p},
  \end{equation}
  where $p$ is the number of parameters (the dimension of $\btheta$) and $\bdelta_{\ML}\equiv\by-\bmM(\btheta_{\ML})$. Under the Null Hypothesis (NH) where $\bdelta_{\ML}$ consists solely of Gaussian noise, the expectation value and variance of $\chisquaredred$ are
  \begin{equation}
    \E[\chisquaredred] = 1 \ \ \text{ and } \ \ \Var[\chisquaredred] = \frac{2}{N-p}. \label{eq:chi-squared-moments}
  \end{equation}
  This is a natural goodness of fit statistic because it has a known distribution under the NH and, disregarding additive normalization constants, it is proportional to the negative log of the likelihood. So, for constant parameter number $p$, minimizing $\chisquaredred$ is identical to maximizing $\mL(\btheta)$.

\subsection{Residual correlations and $\psisquaredred$ statistic}

  The $\chisquaredred$ statistic is only one way of summarizing the proximity of $\bmM(\btheta_{\ML})$ to $\by$ and the Gaussianity of their difference, $\bdelta_{\ML}$. Another way of examining $\bdelta_{\ML}$ is to look at its correlations when it is normalized by the covariance. For the remainder of the paper, the $\ML$ label on $\bdelta$ is neglected for the sake of clarity.

\subsubsection{Definitions} \label{sec:definitions}

The first step in analyzing $\bdelta$ is to form the normalized version $\bDelta\equiv\bC^{-1/2}\bdelta$, which is equivalent to $\Delta_k=\frac{\delta_k}{\sigma_k}$ when the covariance is diagonal. With this definition, $\chisquaredred$ is given by $\frac{\bDelta^T\bDelta}{N-p}$ and thus is determined only by the absolute values, $|\Delta_k|$. Following another thread, we define the correlation vector $\brho$, with its components
  \begin{equation}
    \rho_q \equiv \frac{1}{N-q}\sum_{k=1}^{N-q}\Delta_k\Delta_{k+q} \ \ \ \ \forall q\in\{0,1,\ldots,N-1\} \label{eq:correlation-definition}
  \end{equation}
  being the correlations of the residuals across channels (or, more generally, eigenvectors of the covariance matrix). In the limit where there are vastly more channels than parameters ($N\gg p$), $\rho_0=\chisquaredred$. Through the NH assumption that the components of $\bDelta$ are independent and normally distributed with zero mean and unit variance, it can be calculated that
  \begin{equation}
    \E[\rho_q] = \delta_{q0} \ \ \text{ and } \ \ \Cov[\rho_q,\rho_r] = \delta_{qr}\frac{1+\delta_{q0}}{N-q}\,, \ \ \text{ with } \ \ \delta_{ij}=\begin{cases} 1 & i = j \\ 0 & i \neq j \end{cases}\,,
    \label{eq:nonzero-correlation-moments}
  \end{equation}
  where $\delta_{ij}$ represents the Kronecker delta function only in this equation. While the covariance above indicates that $\rho_q$ and $\rho_r$ are uncorrelated if $q\neq r$, $\rho_q$ and $\rho_r$ are not statistically independent. The standard deviation of the $q^{\text{th}}$ correlation (where $q\neq 0$) is $\sigma_{\rho_q} = \sqrt{\Var[\rho_q]} = \frac{1}{\sqrt{N-q}}$. We then define the reduced psi-squared statistic to be
  \begin{equation}
    \psisquaredred \equiv \frac{1}{N-1}\sum_{q=1}^{N-1}\left(\frac{\rho_q}{\sigma_{\rho_q}}\right)^2. \label{eq:psisquared-definition}
  \end{equation}
  Appendix~\ref{app:practical-computation} contains a straightforward procedure for calculating $\psisquaredred$ from a residual vector.

  \begin{figure}[t!!]
    \centering
    \includegraphics[width=0.6\textwidth]{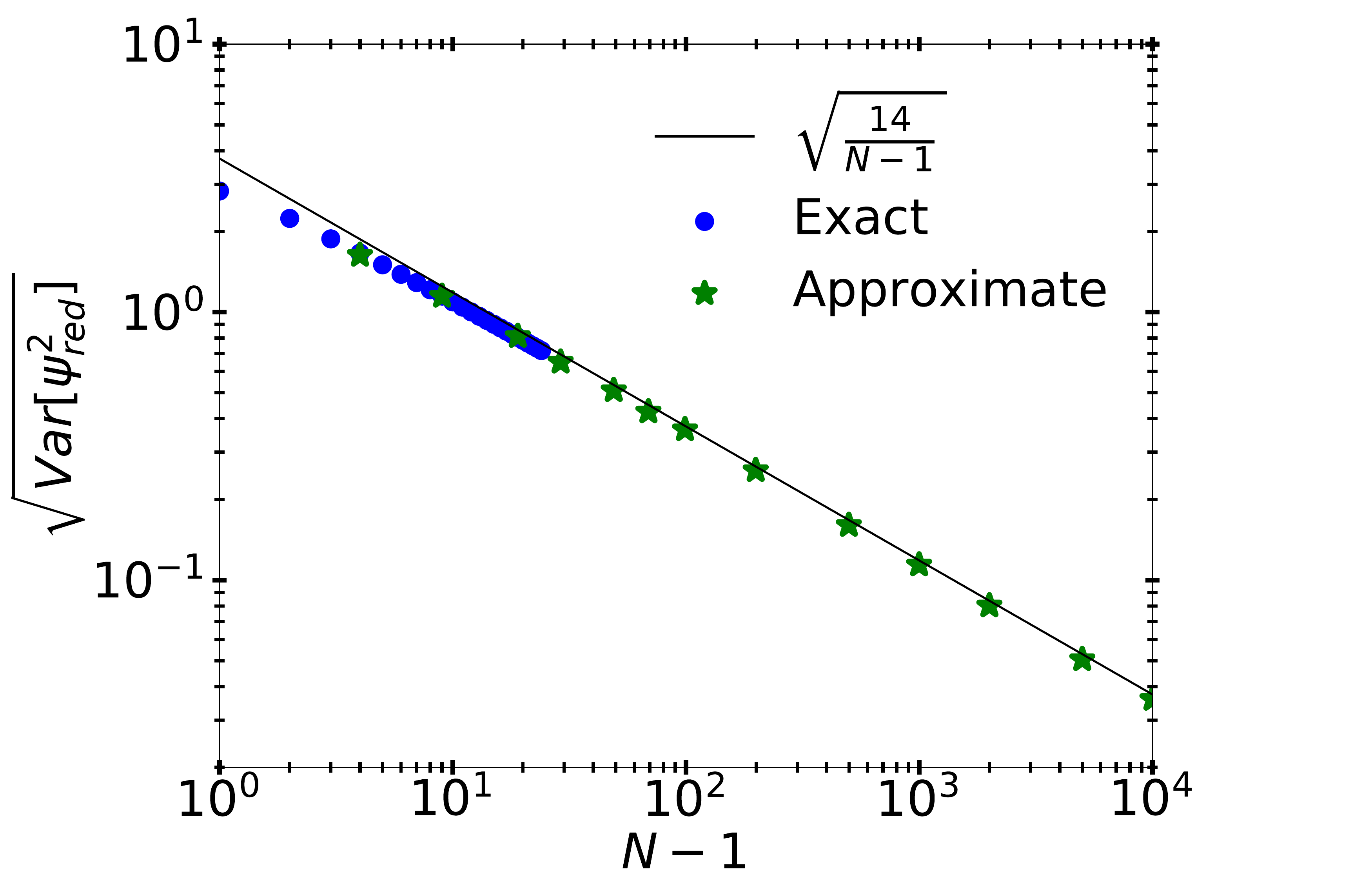}
    \caption{The behavior of the standard deviation of $\psisquaredred$ when the NH is true as a function of sample size. The blue circles show exact values of the standard deviation calculated through Equation~\ref{eq:variance} while the green stars show values of the standard deviation of $\psisquaredred$ approximated through $2\times 10^4$ realizations of noise at a given $N$. The reference line shows the asymptotic $\sqrt{14/(N-1)}$ behavior, which closely mirrors the approximated values across orders of magnitude in $N$.} \label{fig:variance-vs-sample-size}
  \end{figure}

\subsubsection{Null hypothesis distribution} \label{sec:null-hypothesis}

  As is true with $\chisquaredred$, under the NH, the expected value of $\psisquaredred$ is 1. This follows from the fact that, since $\E[\rho_k]=0$, the variance of $\rho_k$ is $\sigma_{\rho_k}^2=\E[{\rho_k}^2]$. The variance of $\psisquaredred$ is difficult to calculate exactly for large $N$, even with computational methods which implement symbolic algebra; but, it is given by
  \begin{equation}
    \Var[\psisquaredred] = \left\{\sum_{\alpha=1}^{N-1} \sum_{\beta=1}^{N-1} \sum_{\gamma=1}^{N-\alpha}\sum_{\delta=1}^{N-\alpha} \sum_{\epsilon=1}^{N-\beta}\sum_{\zeta=1}^{N-\beta} \frac{\E[\Delta_\gamma \Delta_{\gamma+\alpha} \Delta_\delta \Delta_{\delta+\alpha} \Delta_{\epsilon} \Delta_{\epsilon+\beta} \Delta_\zeta\Delta_{\zeta+\beta}]}{(N-1)^2 (N-\alpha) (N-\beta)}\right\} - 1 \label{eq:variance}.
  \end{equation}
  Values of this variance computed using Isserlis' theorem \cite{isserlis-1918} (similar to Wick's theorem) and \texttt{Mathematica}\footnote{\url{http://www.wolfram.com/mathematica/}} for several small sample sizes are shown in Figure~\ref{fig:variance-vs-sample-size} (blue circles). Since $\psisquaredred$ is an average of $N-1$ uncorrelated, zero-mean, unit-variance random variables, it is reasonable to suspect that, asymptotically, $\sigma_{\psisquaredred}\propto (N-1)^{-1/2}$. Indeed, simulations (green stars) that match the exact results at low $N$ show that the formula $\sigma_{\psisquaredred}=\sqrt{14/(N-1)}$ is accurate across many orders of magnitude, leading us to conclude that, for large $N$,
  \begin{equation}
  \E[\psisquaredred] = 1 \ \ \text{ and } \ \ \Var[\psisquaredred] = \frac{14}{N}. \label{eq:psisquaredred-nh-moments}
  \end{equation}
  Thus, the distribution of $\psisquaredred$ is similar to that of $\chisquaredred$ except it is $\sqrt{7}\approx 2.65$ times broader.

\begin{figure}[t!!]
    \centering
    \includegraphics[width=0.32\textwidth]{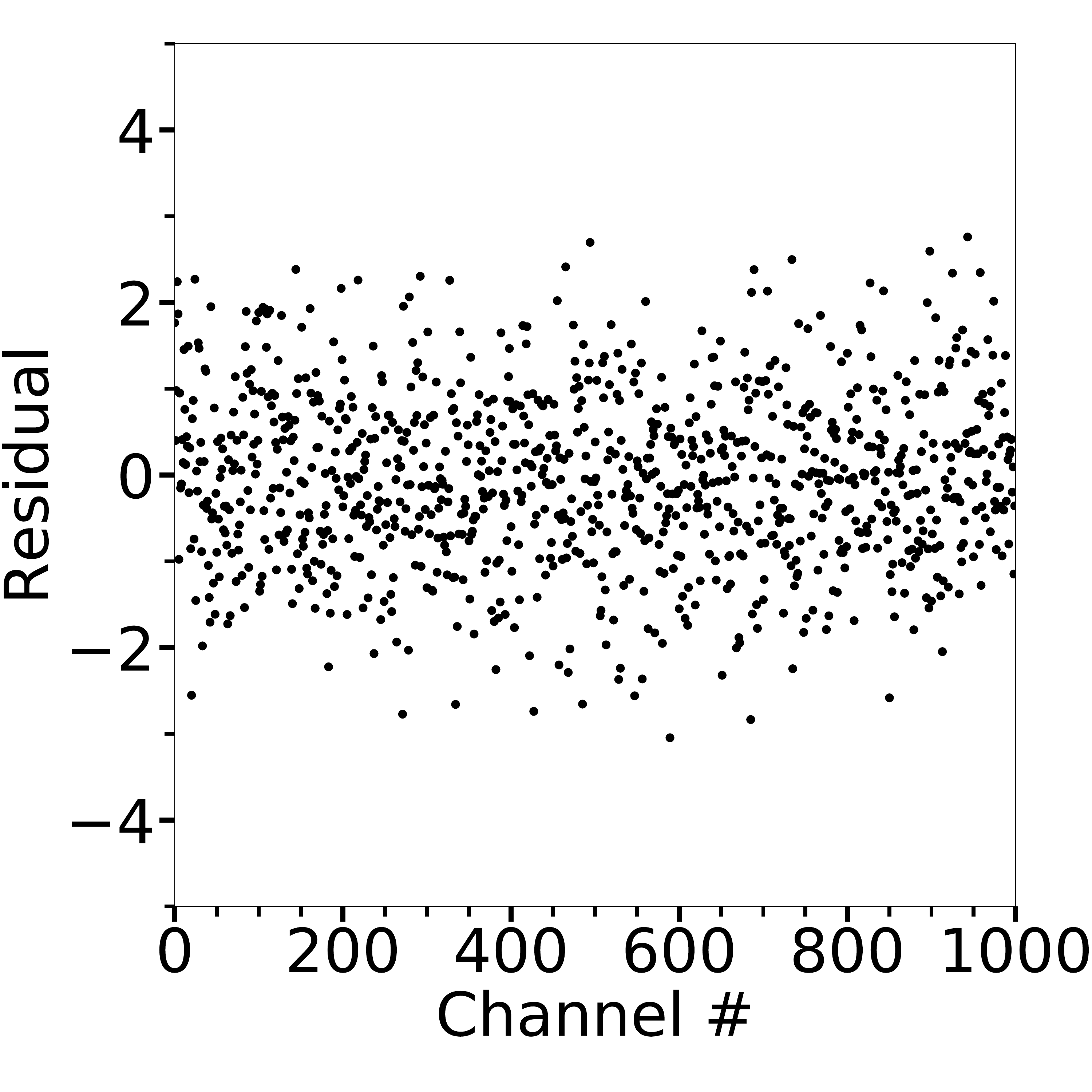}
    \includegraphics[width=0.32\textwidth]{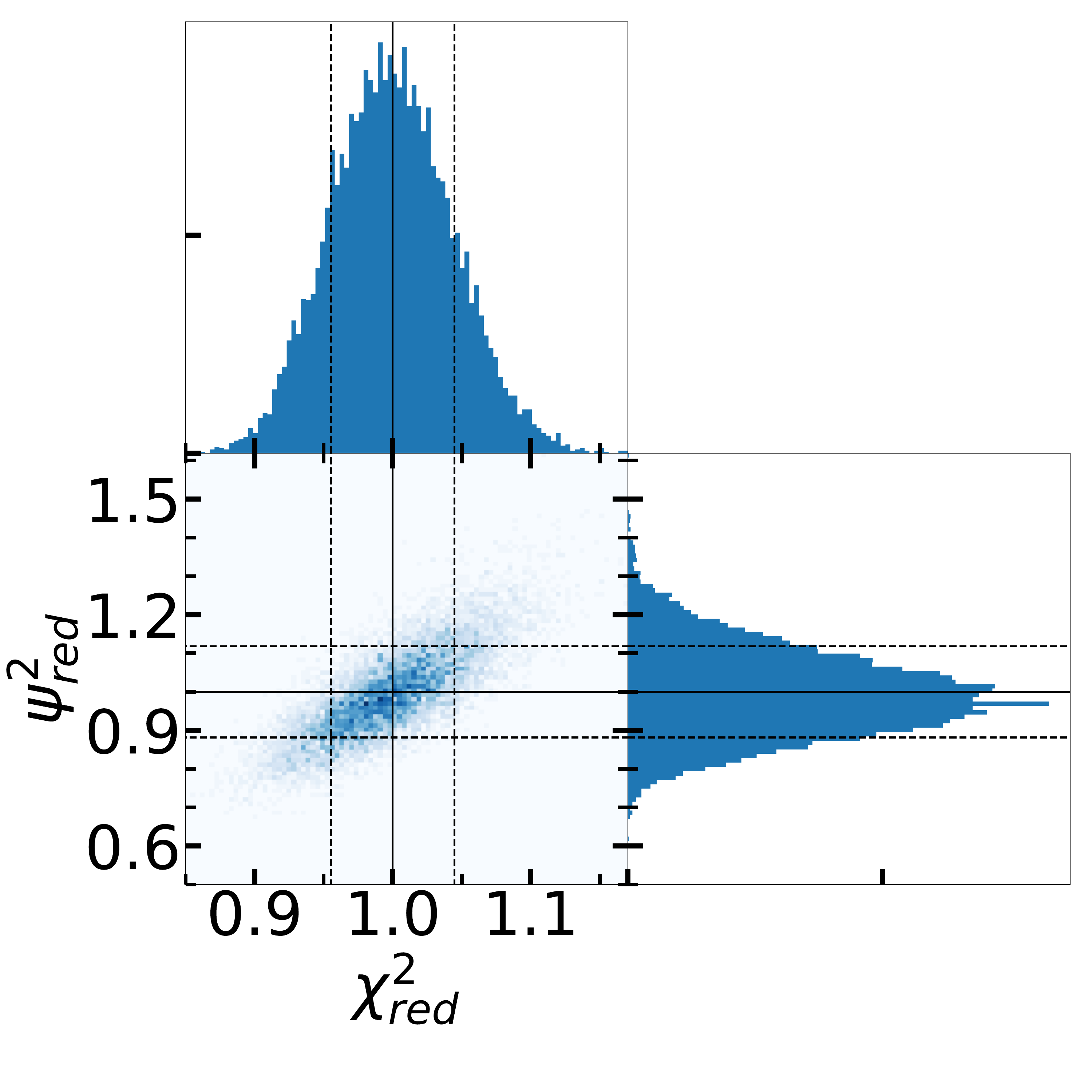}
    \includegraphics[width=0.32\textwidth]{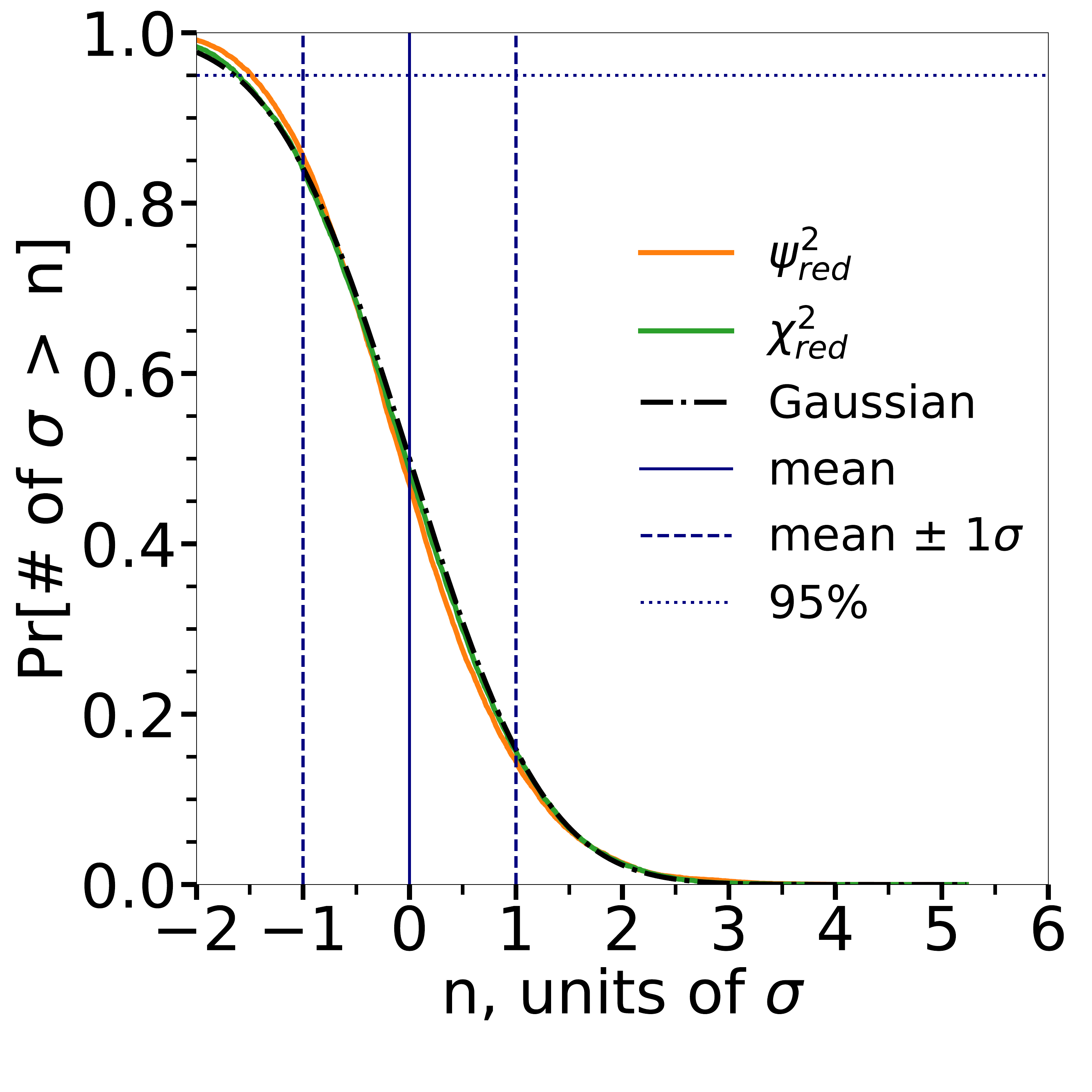}
    \caption{\textit{Left}: Realization of 1000 channels of independent, zero-centered, unit-variance, Gaussian noise. This realization has $\chisquaredred=0.976$ ($\chisquaredred-1=-0.53\,\sigma_{\chisquaredred}$) and $\psisquaredred=0.919$ ($\psisquaredred-1=-0.68\,\sigma_{\psisquaredred}$). \textit{Center}: Joint and marginalized distributions for $\psisquaredred$ and $\chisquaredred$ under the NH where $N=1000$ and $\bDelta$ is unit-variance white noise as is shown in the left panel. The distributions were calculated with $10^4$ different noise realizations. The solid lines mark 1 on each axis while the dashed lines indicate $1\sigma$ deviations from the mean ($1\pm\sqrt{\frac{2}{N}}$ for $\chisquaredred$ and $1\pm\sqrt{\frac{14}{N-1}}$ for $\psisquaredred$). \textit{Right}: The probability that each statistic will be more than $n\sigma$ away from 1 as a function of $n$ under the NH. Solid and dashed blue, vertical lines refer to the center and bounds of the $1\sigma$ confidence interval around the mean as in the center panel. The dash-dot gray line indicates the expected behavior of a standard normal Gaussian variable. The horizontal, dotted line shows 95\% confidence.} \label{fig:null-hypothesis}
  \end{figure}

 The joint distribution and marginal distributions for $\psisquaredred$ and $\chisquaredred$ under the NH for $N=1000$ are shown in Figure~\ref{fig:null-hypothesis}. The estimate of the joint distribution shows that, under the NH where $\bDelta$ is pure standard normal white noise, $\psisquaredred$ is highly correlated with $\chisquaredred$, with a correlation coefficient, $\Corr[\psisquaredred,\chisquaredred]\equiv\frac{\Cov[\psisquaredred,\chisquaredred]}{\sqrt{\Var[\psisquaredred]\ \Var[\chisquaredred]}}$, calculated to be
  \begin{equation}
    \Corr[\psisquaredred,\chisquaredred]_{\text{NH}} \approx 0.8 .
  \end{equation}
  The correlation between $\psisquaredred$ and $\chisquaredred$ in the presence of a feature (i.e. when the NH is false) has an interesting noise level dependence, which is explored in Appendix~\ref{app:correlation-noise-dependence}.

\subsubsection{Correlations and $\psisquaredred$ of simple functions} \label{sec:nonrandom}

\begin{figure}[t!!]
    \centering
    \includegraphics[width=0.96\textwidth]{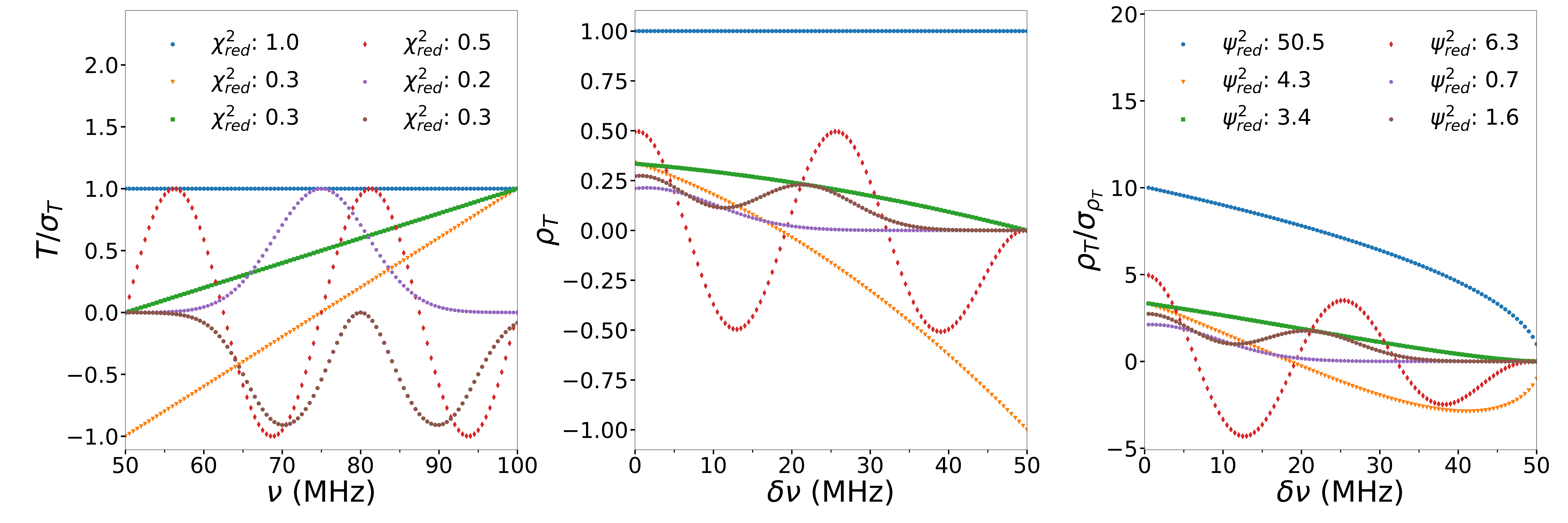}
    \caption{\textit{Left}: Several non-random functions representing fictitious brightness temperature residuals, $T(\nu)$, for a generic 21-cm experiment. The y-axis is normalized to the noise level at each frequency channel. \textit{Center}: The correlation response generated by the curves on the left. Note that $\rho(0)=\chisquaredred$ for each curve. \textit{Right}: A normalized version of the correlation response. $\psisquaredred$ ($\chisquaredred$) is the mean square of the points in the right (left) panel. All plots have $N=101$ points. Increasing $N$ without changing the noise level would lead to an almost identical $\chisquaredred$, whereas $\psisquaredred$ would increase because it is proportional to $N$ (see Equation~\ref{eq:psisquaredred-continuous}).} \label{fig:nonrandom-correlations}
  \end{figure}

  The principle purpose of a goodness-of-fit test is to determine whether after the fit there are remaining non-noise like components in the data. Therefore, it is useful to know how the correlations $\rho_q$, on which $\psisquaredred$ depends, behave when $\bDelta$ is not noise-like. As points get more and more dense for large $N$, when $\bDelta$ follows an underlying curve, $f(\nu)$, we can write a continuous analog of the vector $\brho$ in the form of an integral,
  \begin{equation}
    \rho(\varepsilon) = \frac{1}{\Delta\nu-\varepsilon}\int_{\nu_{\text{min}}}^{\nu_{\text{max}}-\varepsilon}f(\nu)\ f\left(\nu+\varepsilon\right)\ d\nu \ \ \text{ where } \ \ \Delta\nu\equiv\nu_{\text{max}}-\nu_{\text{min}}. \label{eq:correlation-continuous}
  \end{equation}
  Note that, as in the discrete case, $\rho(0)=\chisquaredred$. Using $\psisquaredred=\int_0^{\Delta\nu}\frac{d\varepsilon}{\Delta\nu}\left[\frac{\rho(\varepsilon)}{\sigma_\rho(\varepsilon)}\right]^2$, where $\sigma^2_\rho(\varepsilon) = \frac{1}{N}\frac{1}{1-\frac{\varepsilon}{\Delta\nu}}$ is the expected variance of the correlation under the NH, we find that a function $f(\nu)$ induces a $\psisquaredred$ of
  \begin{equation}
    \psisquaredred \approx \frac{N}{(\Delta\nu)^2}\int_0^1\frac{1}{1-\xi}\left[\int_{\nu_{\text{min}}}^{\nu_{\text{min}} + (1-\xi)\Delta\nu} f(\nu)\ f(\nu+\xi\ \Delta\nu)\ d\nu\right]^2\ d\xi. \label{eq:psisquaredred-continuous}
  \end{equation}
  It is clear that scaling $f(\nu)$ by a constant $c$ will result in a scaling of $\rho(\varepsilon)$ by $c^2$ and a scaling of $\psisquaredred$ by $c^4$. Values of the correlation $\rho$, its normalized counterpart, and $\psisquaredred$ are shown for various functions in Figure~\ref{fig:nonrandom-correlations}. Analytical formulae for the $\psisquaredred$ of different features are given in Appendix~\ref{app:tables-of-values}.

\subsubsection{Sensitivity to non-random components in the presence of noise} \label{sec:random-and-nonrandom}

  In general, fits produce residuals which are combinations of random, noise-like components and non-random components. Consider a residual vector, $\Delta_k$, composed of a non-random vector $\mu_k$ and a standard normal noise vector $n_k$ such that $\Delta_k=\mu_k+n_k$. Defining $\overline{\rho_q}\equiv\E[\rho_q]$ and using Equation~\ref{eq:correlation-definition}, for $q>0$ we find
  \begin{equation}
    \overline{\rho_q}=\frac{1}{N-q}\sum_{k=1}^{N-q}\mu_k\mu_{k+q}.
  \end{equation}
  By defining $\overline{\psisquaredred}\equiv\frac{1}{N-1}\sum_{q=1}^{N-1}(N-q){\overline{\rho_q}}^2$ (if $\mu_k=f(\nu_k)$, this is approximately given by Equation~\ref{eq:psisquaredred-continuous}), we obtain
  \begin{equation}
    \E[\psisquaredred] = 1 + \overline{\psisquaredred} + \frac{1}{N-1}\sum_{k=1}^N{\mu_k}^2(2H_{N-1}-H_{k-1}-H_{N-k}) + \frac{2}{N-1}\sum_{q=1}^{\lfloor(N-1)/2\rfloor} \frac{N-2q}{N-q}\overline{\rho_{2q}} \label{eq:full-psisquaredred}
  \end{equation}
  where $H_n=\sum_{\alpha=1}^n\frac{1}{\alpha}$ is the $n^{\text{th}}$ harmonic number. The first term, 1, comes only from the noise, the second term, $\overline{\psisquaredred}$, comes only from the non-random component and corresponds to the $\psisquaredred$ statistic computed in Section~\ref{sec:nonrandom} (see, e.g., Figure~\ref{fig:nonrandom-correlations}), and the two sums come from the interaction of the noise with the non-random component. Using the general inequality $2ab\ge -a^2-b^2$, Equation~\ref{eq:full-psisquaredred} can also be cast into a useful inequality,\footnote{\label{footnote:stronger-inequality}The stronger inequality $\E[\psisquaredred] \ge 1 + \overline{\psisquaredred} + \frac{1}{N-1}\sum_{n=1}^N{\mu_n}^2(H_{\lceil(N+n)/2\rceil-1}+H_{N-\lfloor(n+1)/2\rfloor}-H_{n-1}-H_{N-n})$ also holds.}
  \begin{equation}
    \E[\psisquaredred] \ge 1 + \overline{\psisquaredred} + \varphi\ \overline{\chisquaredred}, \label{eq:psisquaredred-inequality}
  \end{equation}
  where $\varphi=\ln{\left(\frac{9}{4}\right)}\approx 0.811$. Interestingly, the last term indicates that the mean of the distribution of $\psisquaredred$ is nearly as sensitive to single channel spikes (i.e. individual extreme outliers) as the mean of the distribution of $\chisquaredred$, even though these $\mu_k$'s have $\overline{\psisquaredred}=0$ (see Appendix~\ref{app:spike-sensitivity} for details). Equation~\ref{eq:psisquaredred-inequality} is similar to the corresponding equation for $\chisquaredred$, $\E[\chisquaredred]=1+\overline{\chisquaredred}$, except that in almost all cases, the relation of Equation~\ref{eq:psisquaredred-inequality} is far from equality because the non-random correlations amplify the variance of the random correlations, an effect absent in the $\chisquaredred$ case.

  \begin{figure}[t!!]
    \centering
    \includegraphics[width=0.32\textwidth]{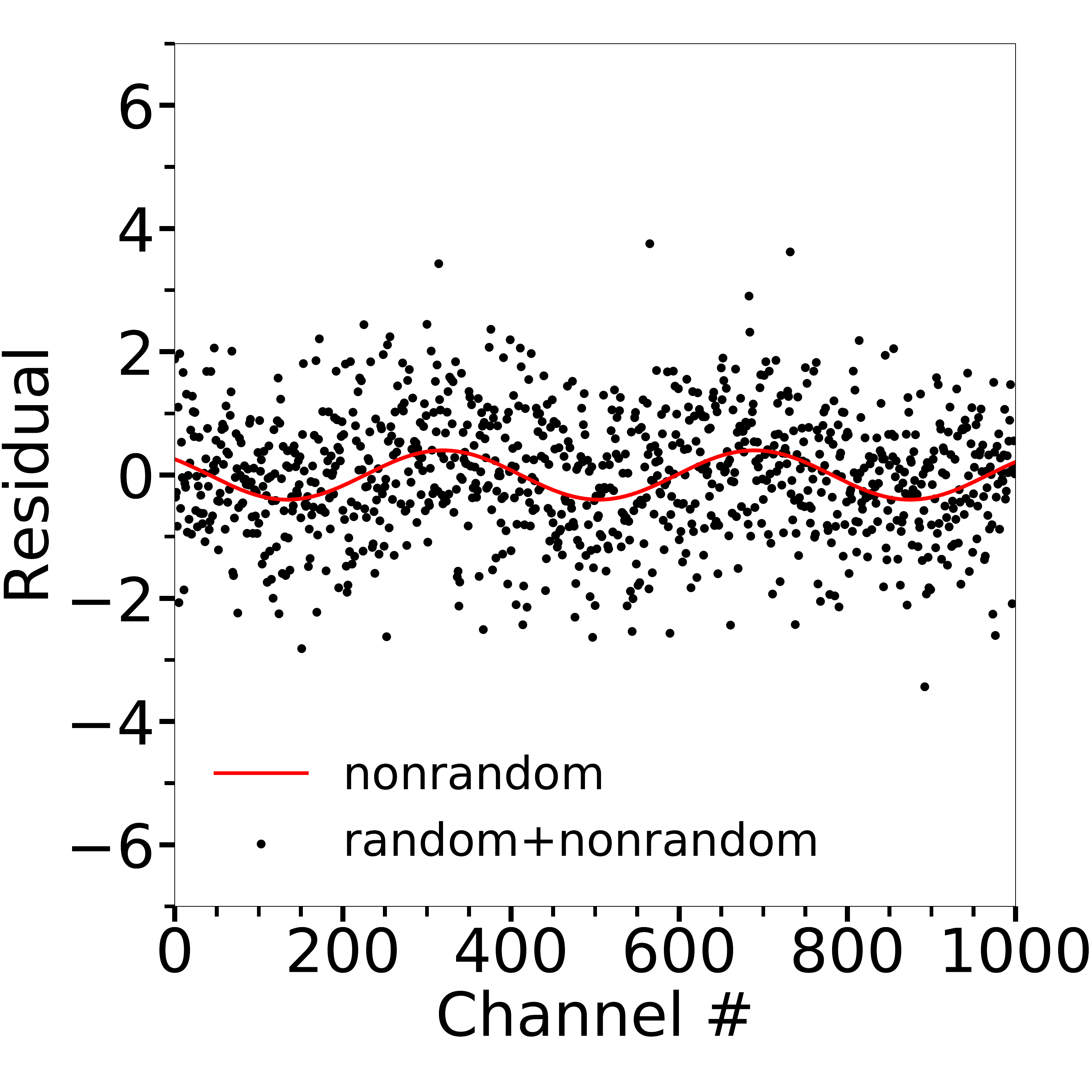}
    \includegraphics[width=0.32\textwidth]{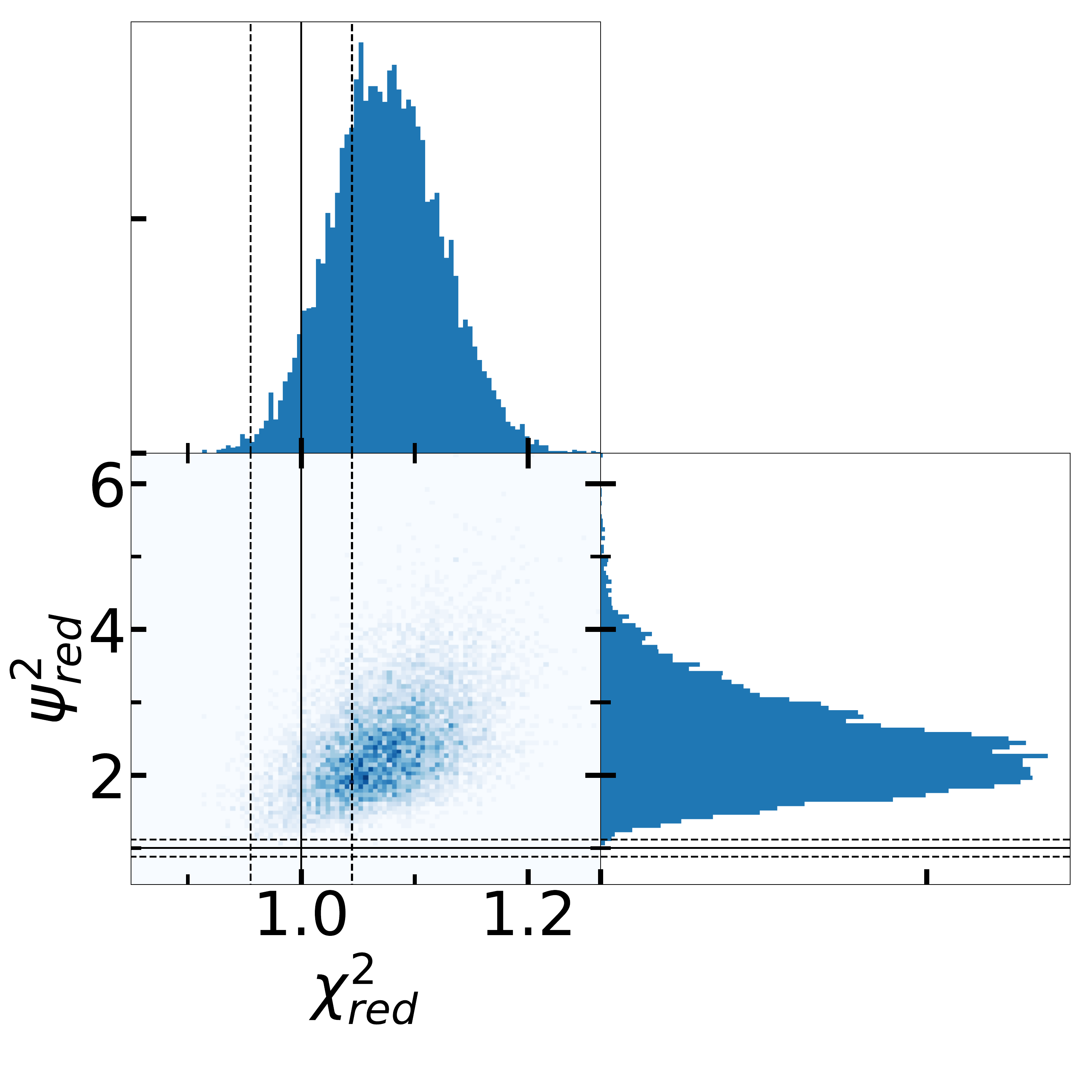}
    \includegraphics[width=0.32\textwidth]{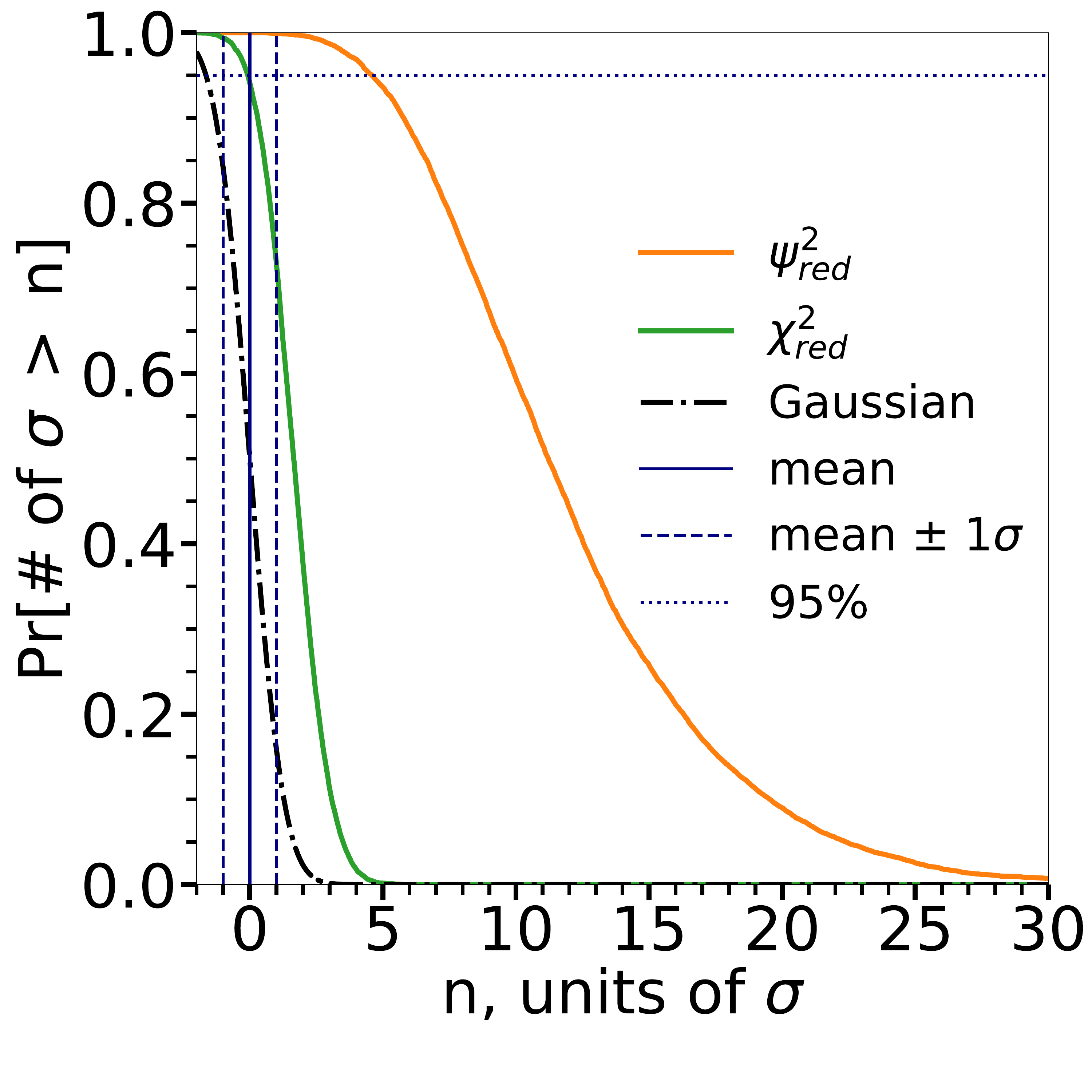}
    \caption{Same as Figure~\ref{fig:null-hypothesis} except all curves used to compute the two statistics consist of the sum of random noise and the non-random sine wave shown in red in the left panel. The realization shown in the left panel has $\chisquaredred=1.005$ ($\chisquaredred-1=0.1\,\sigma_{\chisquaredred}$) and $\psisquaredred=1.283$ ($\psisquaredred-1=2.4\,\sigma_{\psisquaredred}$). The correlation coefficient of $\psisquaredred$ and $\chisquaredred$ is about 0.47 in this case. The right panel shows that $\gtrsim 98\%$ of realizations satisfy $\psisquaredred>1+3\,\sigma_\psisquaredred$, whereas $\lesssim 11\%$ of realizations satisfy $\chisquaredred>1+3\,\sigma_\chisquaredred$. Hence, $\psisquaredred$ is significantly more sensitive to this non-random component than $\chisquaredred$.} \label{fig:nonrandom-sine-wave}
  \end{figure}
  
  \subsubsection{Examples inspired by recent 21-cm results}
  
  For non-random components which are extended across channel space, $\psisquaredred$ is generally more sensitive than $\chisquaredred$. Figure~\ref{fig:nonrandom-sine-wave} shows the distribution of values of $\psisquaredred$ and $\chisquaredred$ induced by a non-random sine wave added into the noise in each simulated realization of $\bDelta$. It was suggested by \cite{hills-2018} that a similar sinusoidal ripple, along with a foreground model, could explain the publicly released EDGES spectrum.
  
\begin{figure}[t!!]
    \centering
    \includegraphics[width=0.32\textwidth]{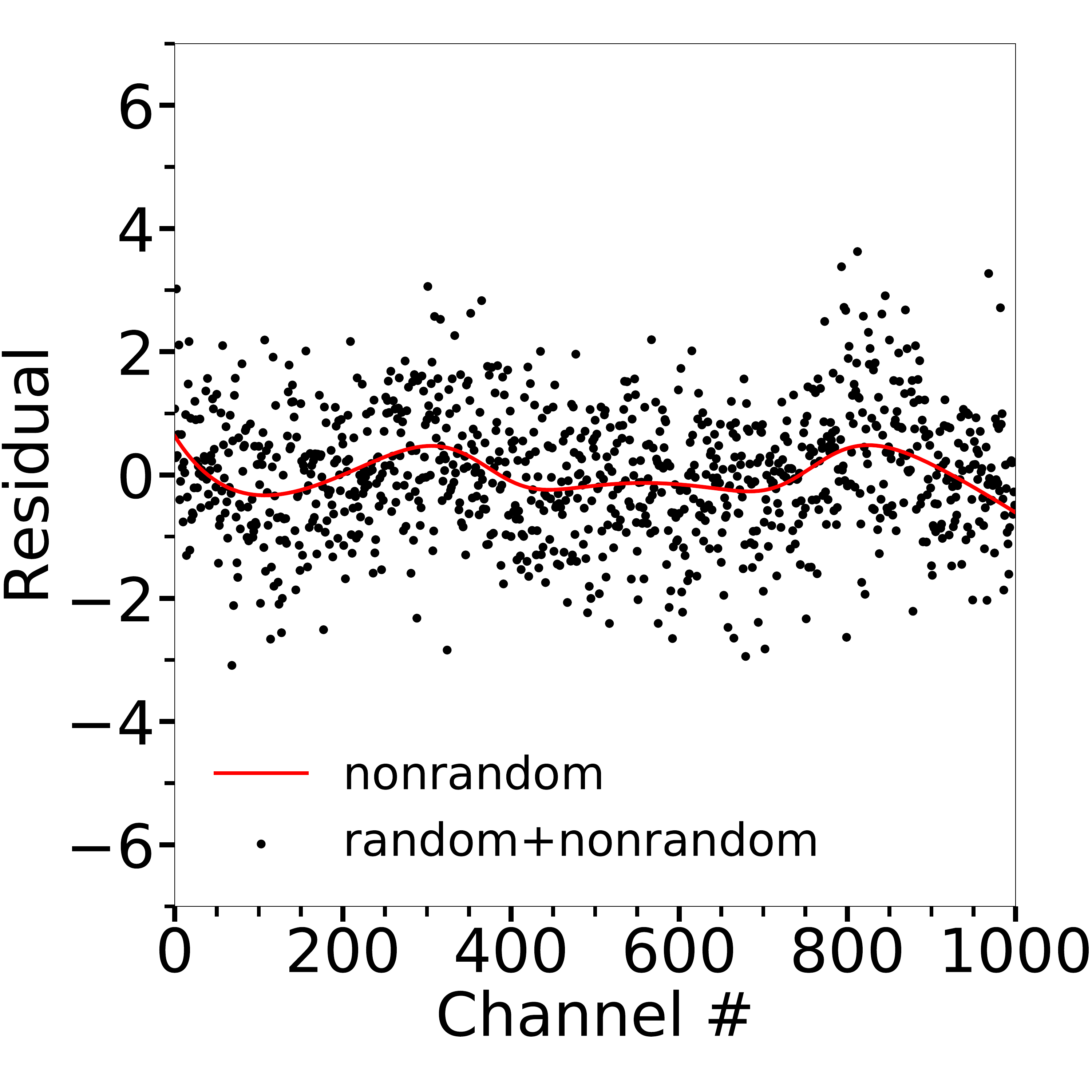}
    \includegraphics[width=0.32\textwidth]{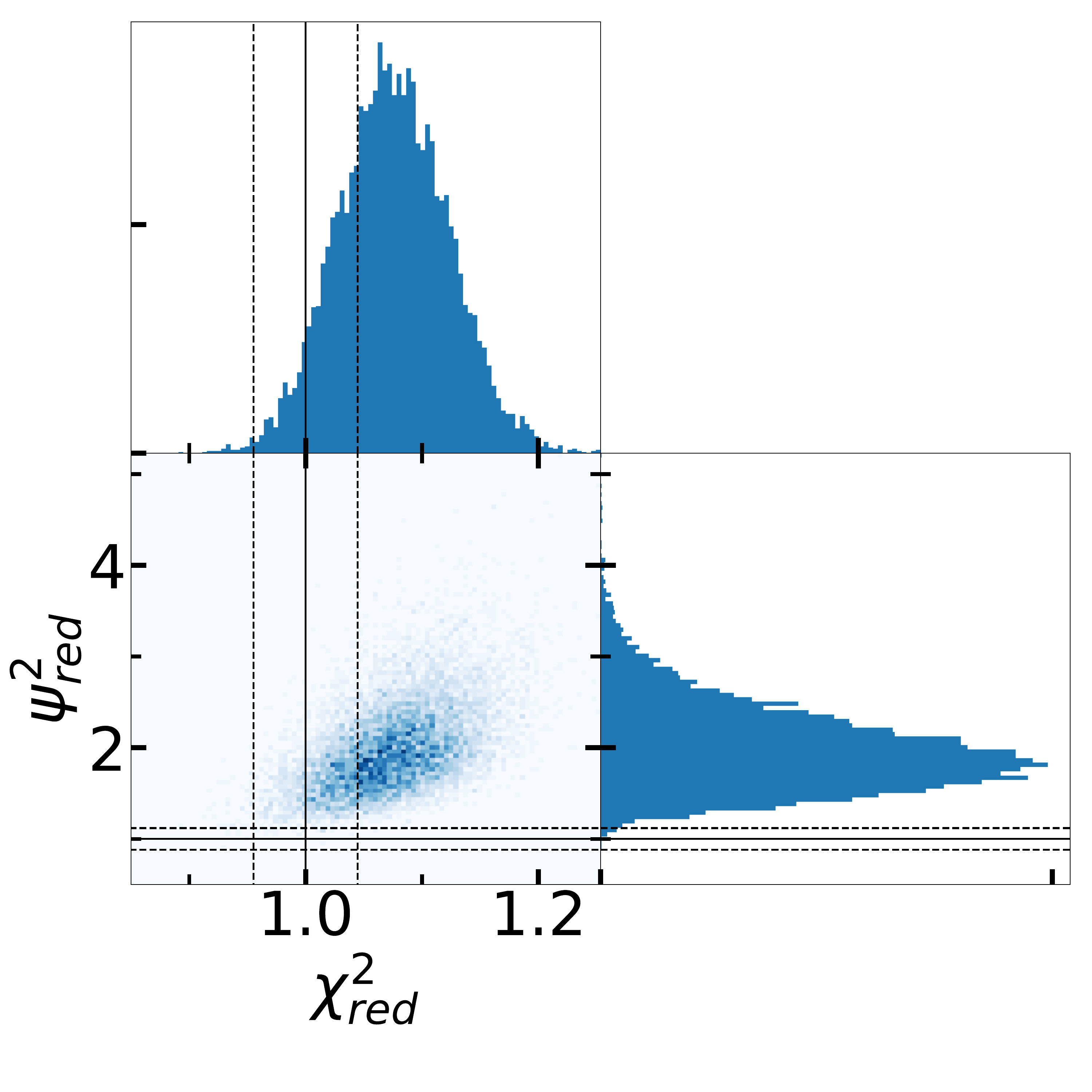}
    \includegraphics[width=0.32\textwidth]{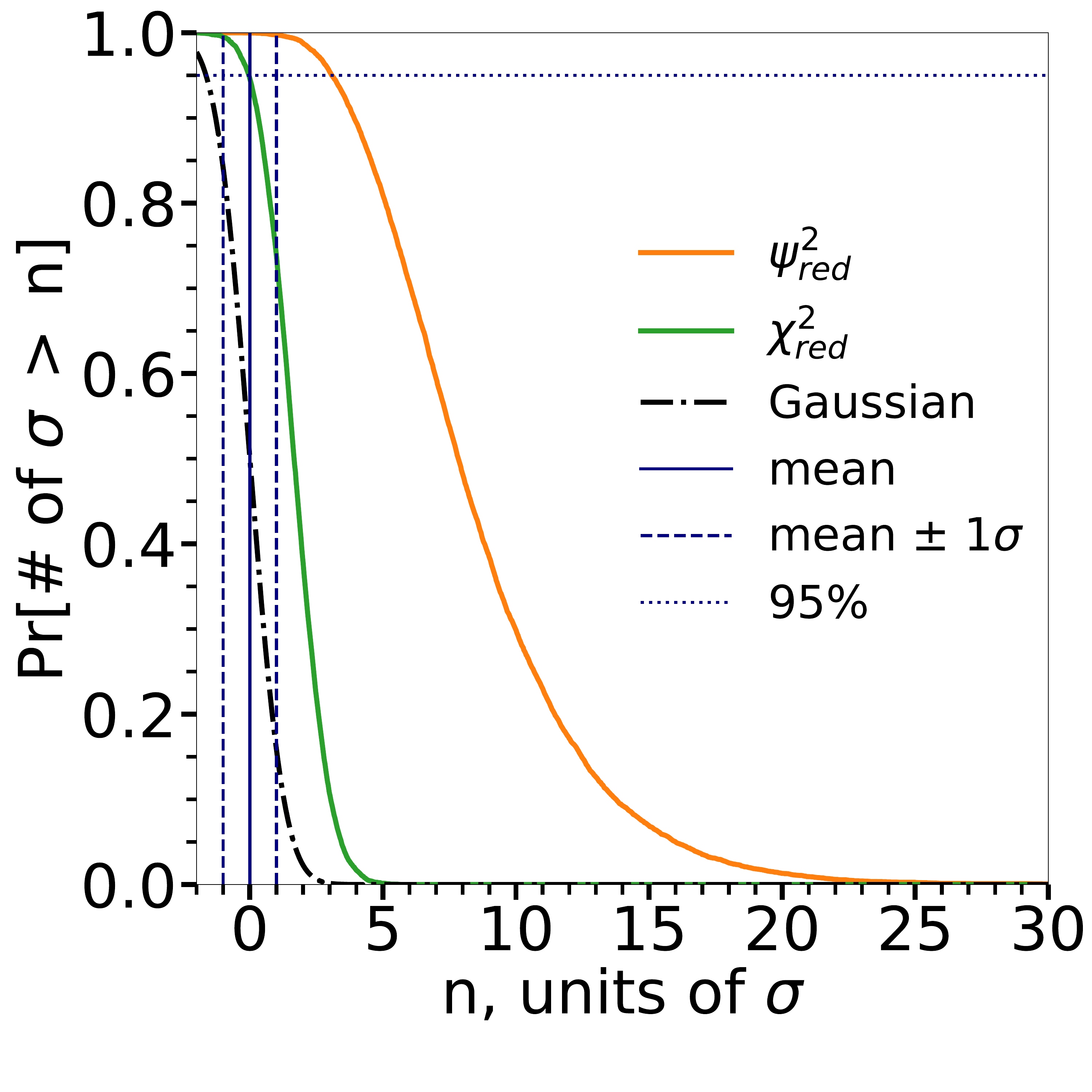}
    \caption{Same as Figure~\ref{fig:null-hypothesis} except that all curves have a non-random curve (shown in red in left panel) added to them (see text for details). The realization shown has $\chisquaredred=1.07$ ($\chisquaredred-1=1.47\,\sigma_{\chisquaredred}$) and $\psisquaredred=1.93$ ($\psisquaredred-1=7.85\,\sigma_{\psisquaredred}$). The correlation between $\chisquaredred$ and $\psisquaredred$ is 0.51. Here, as in the case of Figure~\ref{fig:nonrandom-sine-wave}, $\psisquaredred$ is more sensitive than $\chisquaredred$.} \label{fig:nonrandom-ideal-residual}
  \end{figure}

  Figure~\ref{fig:nonrandom-ideal-residual} shows the distributions of $\psisquaredred$ and $\chisquaredred$ when a different non-random component is added in. In this case, the non-random component is the residual when a flattened Gaussian, as introduced in \cite{bowman-2018}, of the form
  \begin{align}
    T_{\text{sys}} = A\left(\frac{1-e^{-\tau e^B}}{1-e^{-\tau}}\right) \ \ \text{ where } \ \ B = \frac{4(\nu-\mu)^2}{w^2}\ln{\left[-\frac{1}{\tau}\ln{\left(\frac{1+e^{-\tau}}{2}\right)}\right]} \label{eq:flattened-gaussian}
  \end{align}
  with $A=-600$ mK, $\mu=78$ MHz, $\tau=7$, and $w=20$ MHz is fit with a foreground model consisting of a power law with spectral index $-2.5$ times a six-term polynomial (see Equation~\ref{eq:power-law-times-polynomial}). This curve is similar to the dashed lines of `Extended Data Figure 8' in \cite{bowman-2018}, which recently reported finding such a flattened Gaussian in the sky-averaged spectrum of low-band EDGES data. Therefore, given that the noise level in that case is lower than the amplitude of the non-random component, fitting the data with only a foreground model should lead to a fit with residuals at least as poor as those shown in the left panel of Figure~\ref{fig:nonrandom-ideal-residual}. Importantly, from the center and right panels of this figure we see then that having a noise level estimate for the data --- which is not presented in \cite{bowman-2018} --- should allow us to calculate $\psisquaredred$ for a strong statistical test of whether there is a flattened Gaussian-like feature in the data as claimed.
  
\subsubsection{Hypothesis testing with $\psisquaredred$} \label{sec:hypothesis-test}

  We proceed here to define a formal hypothesis test which indicates if $\bDelta$ has non-noise-like components above a specific confidence level. We do so by assuming the NH --- that $\bDelta$ is pure white noise --- and a significance level $\alpha$ (which corresponds to a confidence of $1-\alpha$; see the $y$ coordinate in the right panel of Figure~\ref{fig:null-hypothesis}), and calculating the constant $\zeta$ such that $\Pr[\psisquaredred > \zeta] = \alpha$. Since $\psisquaredred$ is approximately Gaussian with mean 1 and variance $\frac{14}{N}$, $\zeta$ can be calculated using the error function. We find that
  \begin{align}
    \Pr\left[\psisquaredred>1+\sqrt{\frac{28}{N}}\ \text{erf}^{-1}{(1-2\alpha)}\right] = \alpha. \label{eq:hypothesis-test}
  \end{align}
  Choosing a significance level of $\alpha=10^{-3}$, Equation~\ref{eq:hypothesis-test} can be written as
  \begin{align}
    \Pr\left[\frac{\psisquaredred-1}{\sigma_{\psisquaredred}}>3.1\right]=10^{-3}. \label{eq:concrete-hypothesis-test}
  \end{align}
  This equation also holds for $\chisquaredred$ and $\sigma_{\chisquaredred}$, except that the $\chisquaredred$ version is less sensitive to low-level wide-band features than the $\psisquaredred$ version, as can be seen in the fact that the $\chisquaredred$ curves (green) in the right panels of Figures~\ref{fig:nonrandom-sine-wave}~and~\ref{fig:nonrandom-ideal-residual} are shifted to the right less than their $\psisquaredred$ counterparts (orange). Equation~\ref{eq:concrete-hypothesis-test} means that if $\psisquaredred-1>3.1\,\sigma_{\psisquaredred}$, then we can conclude, with 99.9\% confidence, that the residual $\bDelta$ is not made purely of random white noise. This could mean either that the estimate of the channel covariance $\bC$ used to normalize the residuals is incorrect or that the model is insufficient to fit all non-noise-like structure in the data.
  
\begin{figure}[t!!]
    \centering
    \includegraphics[width=0.32\textwidth]{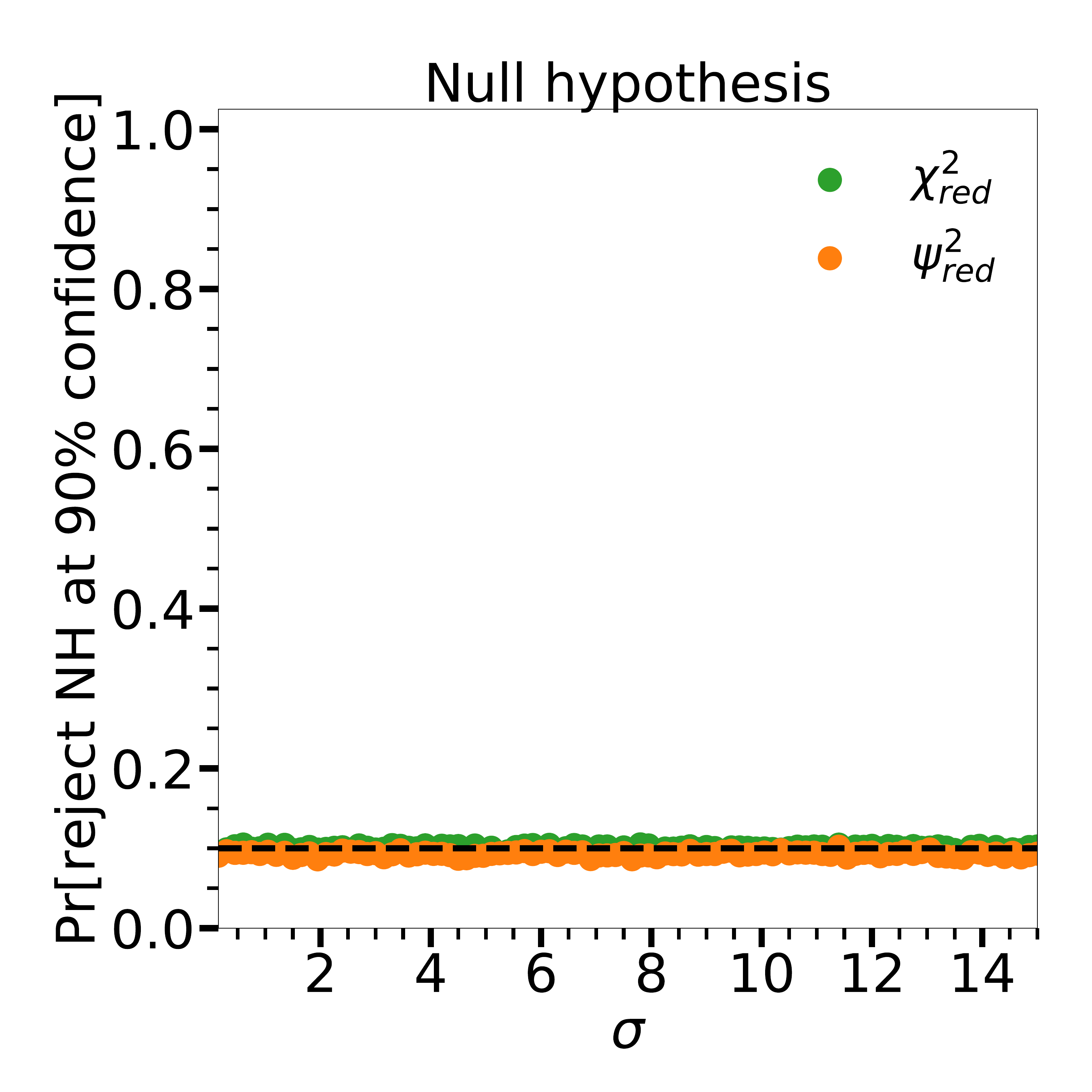}
    \includegraphics[width=0.32\textwidth]{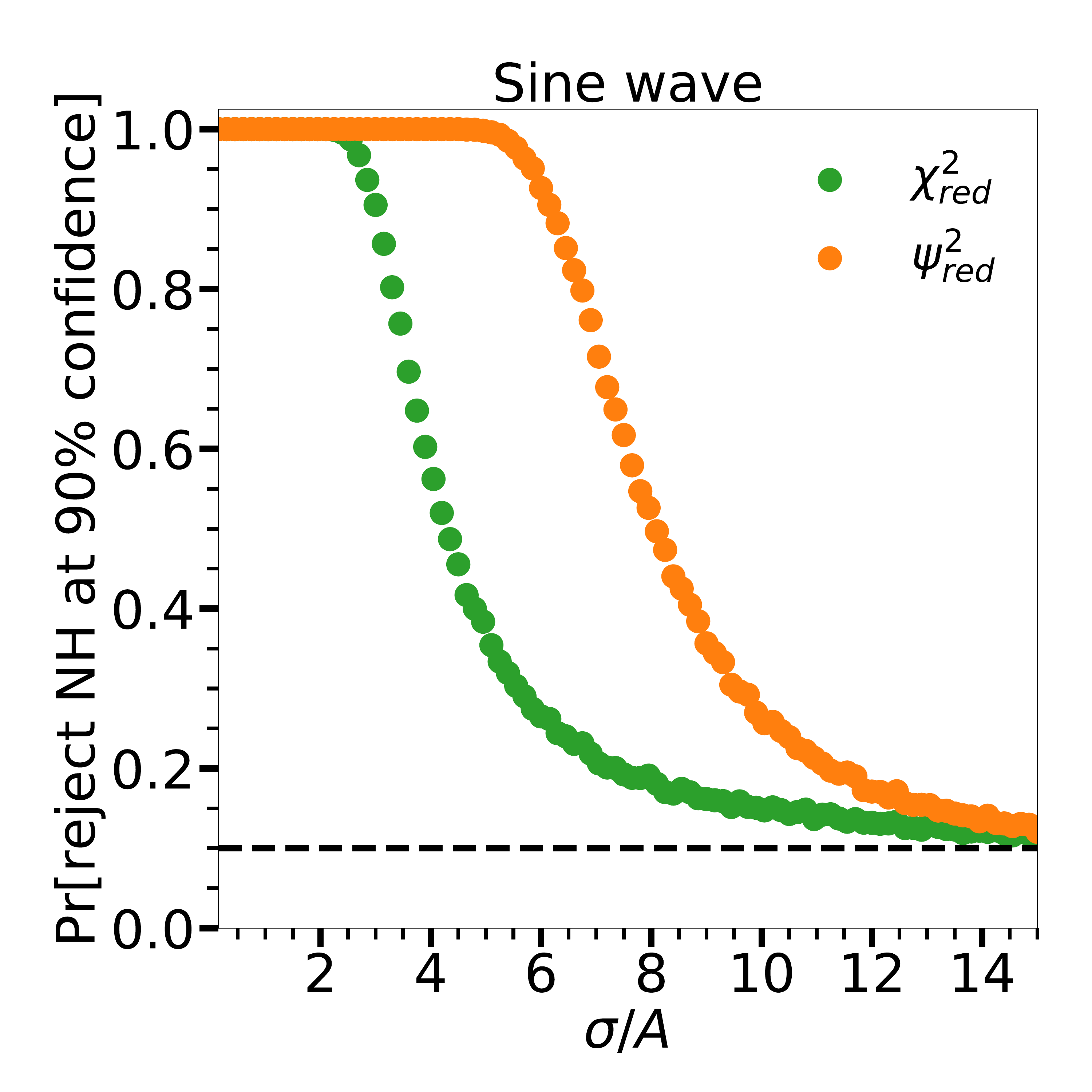}
    \includegraphics[width=0.32\textwidth]{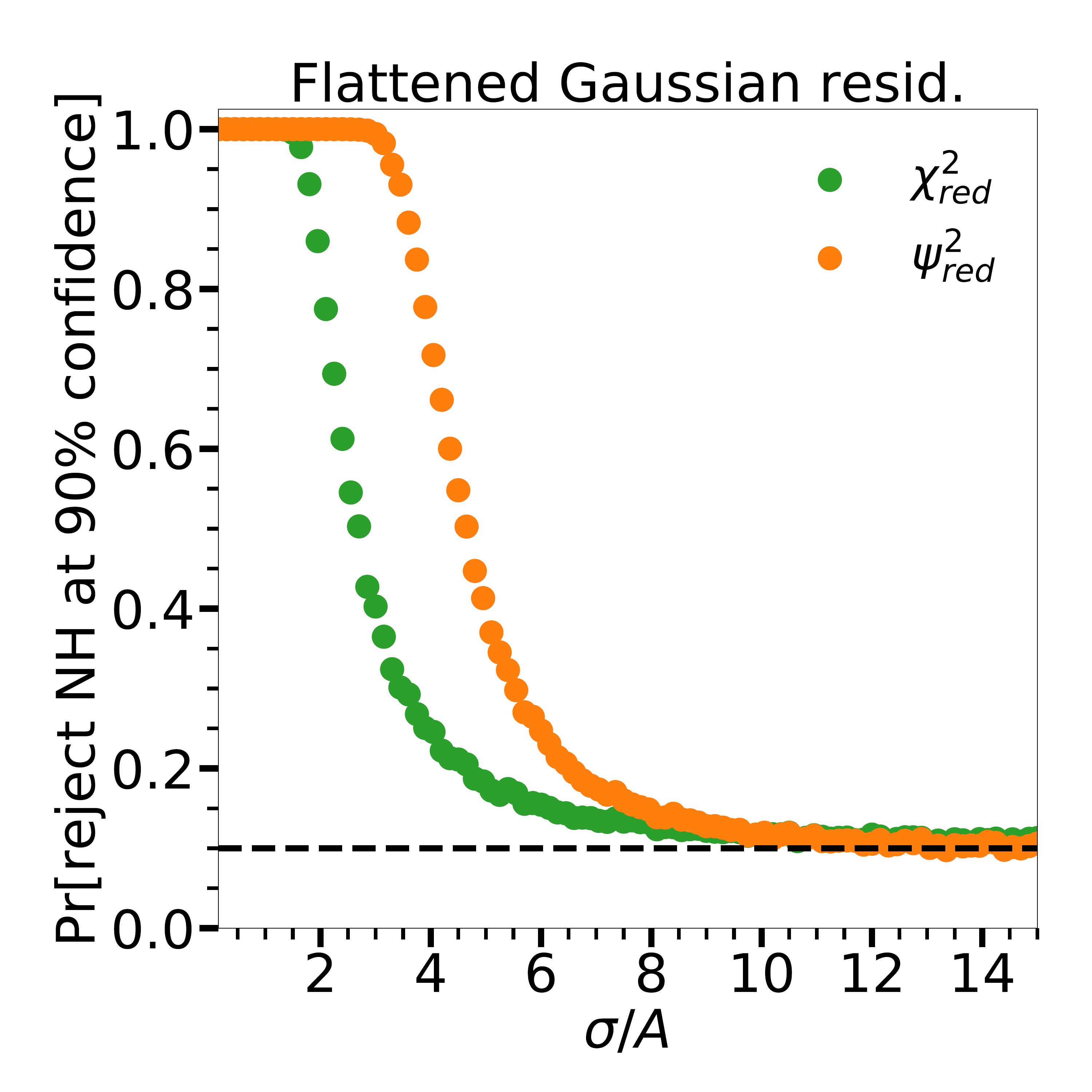}
    \caption{Plots of the probability of rejecting the NH at 90\% confidence for both $\psisquaredred$ and $\chisquaredred$ as a function of the noise level for the three features from Figures~\ref{fig:null-hypothesis},~\ref{fig:nonrandom-sine-wave}~and~\ref{fig:nonrandom-ideal-residual} (zero for the NH case, and the corresponding red lines from Figures~\ref{fig:nonrandom-sine-wave}~and~\ref{fig:nonrandom-ideal-residual} for the sine wave and flattened Gaussian cases). Each point is calculated from $10^4$ realizations of $5001$ channels, with noise at the given level added to the feature in question. The points in each curve span $100$ different noise levels. The x-axes of the center and right panels are normalized by the maximum absolute value of the given feature, $A$. The dashed black line in all of the panels is the confidence floor, which is $10\%$ for a confidence of $90\%$.}
    \label{fig:rejection-probability-vs-noise-level}
  \end{figure}

  It can also be seen from the cumulative distribution functions in the right panels of Figures~\ref{fig:nonrandom-sine-wave}~and~\ref{fig:nonrandom-ideal-residual} that the probability of $\psisquaredred$ being more than $3.1\,\sigma_\psisquaredred$ away from 1 is greater than 95\% in the case of both figures, whereas the same probability for $\chisquaredred$ is only about 10\%. Another way of visualizing the relative utility of $\psisquaredred$ and $\chisquaredred$ is to examine the probability of rejecting the NH at a specific confidence as a function of the noise level. For a confidence level of $90\%$, this probability is shown in Figure~\ref{fig:rejection-probability-vs-noise-level} for the features (or lack thereof) which generated Figures~\ref{fig:null-hypothesis},~\ref{fig:nonrandom-sine-wave},~and~\ref{fig:nonrandom-ideal-residual}. The left panel shows that, as expected for a confidence level of $90\%$, the NH is rejected $\sim 10\%$ of the time even when it is true. The other panels show that, for the features shown by the red lines in Figures~\ref{fig:nonrandom-sine-wave}~and~\ref{fig:nonrandom-ideal-residual}, $\psisquaredred$ can be used to reject the NH at $90\%$ confidence out to a larger noise level than can $\chisquaredred$.
  
  Clearly, $\psisquaredred$ is a powerful statistic to ascertain if large scale feaures remain in fit residuals. Such large scale features are common in wideband applications like 21-cm cosmology, especially when the smooth foreground features are not fit well down to the requisite level.

\section{Application to 21-cm cosmology} \label{sec:21-cm-cosmology}

  This section concerns the distribution of $\psisquaredred$ and $\chisquaredred$ when they are computed using residuals of fits performed in the presence of noise, as opposed to their distributions when noise is added to a non-random curve. As a working example, we choose the curves that are fit to be similar to data components from experiments which are attempting to measure the sky-averaged, highly-redshifted 21-cm spectrum of neutral hydrogen.
  
  In general, global 21-cm signal experiments measure data which are the combination of four components: a) galactic and extra-galactic foregrounds weighted by the antenna beam, b) the global 21-cm signal, c) instrument systematics, and d) Gaussian noise that follows the radiometer equation, describing noise proportional to the data itself \cite{condon-2016}. With a single set of simulated data vectors that include these four components, we perform three fits with models of differing sufficiency to show the reaction of $\psisquaredred$ and $\chisquaredred$ to residuals encountered after an actual fitting. The first fit includes only a foreground model, the second fit includes models of the foreground and instrument systematic, and the final fit sufficiently models all three non-random components in the data.

\subsection{Models} \label{sec:21-cm-models}
  
  For the purpose of fitting the foregrounds in this paper, we use a power law times polynomial model (this model is used in \cite{bowman-2018}),
  \begin{align}
    T_{\text{fg}} = \left(\frac{\nu}{\nu_0}\right)^{-2.5}\sum_{k=1}^{N_{\text{terms}}}a_k\left(\frac{\nu}{\nu_0}\right)^{k-1}. \label{eq:power-law-times-polynomial}
  \end{align}
  Because the foregrounds that appear in the data are a complex sum of individual sources weighted by the antenna beam, models which are specific to the given antenna beam and sky position are preferred (see \cite{tauscher-2018} for a method of creating such a model); but, the polynomials of Equation~\ref{eq:power-law-times-polynomial} serve as a useful generic model to showcase typical distributions of $\psisquaredred$ in common fitting applications in 21-cm cosmology. It is important to note that the polynomial model is only effective if the true beam-weighted foreground in the data is well-fit by such a polynomial, an implicit assumption which we can ensure is true in our simulated exercise, but may be dubious when analyzing real data. We compute the 21-cm signal with a phenomenological model that represents relevant astrophysical quantities through parameters in hyperbolic tangent functions (see \cite{harker-2016} for details on this model). As an instrument systematic for our exercise, we choose a Lorentzian, suggested by Bradley et al. (in prep.) to approximately model one\footnote{While only one Lorentzian is included in our example, this would naturally be accompanied by others, as detailed in Bradley et al. (in prep.), that can be within or outside of the frequency range considered.} absorptive ground plane resonance,
  \begin{align}
    T_{\text{sys}} = \frac{A}{1+\left(\frac{\nu-\mu}{\sigma}\right)^2}. \label{eq:lorentzian}
  \end{align}

  \begin{figure}[t!!]
    \centering
    \includegraphics[width=0.32\textwidth]{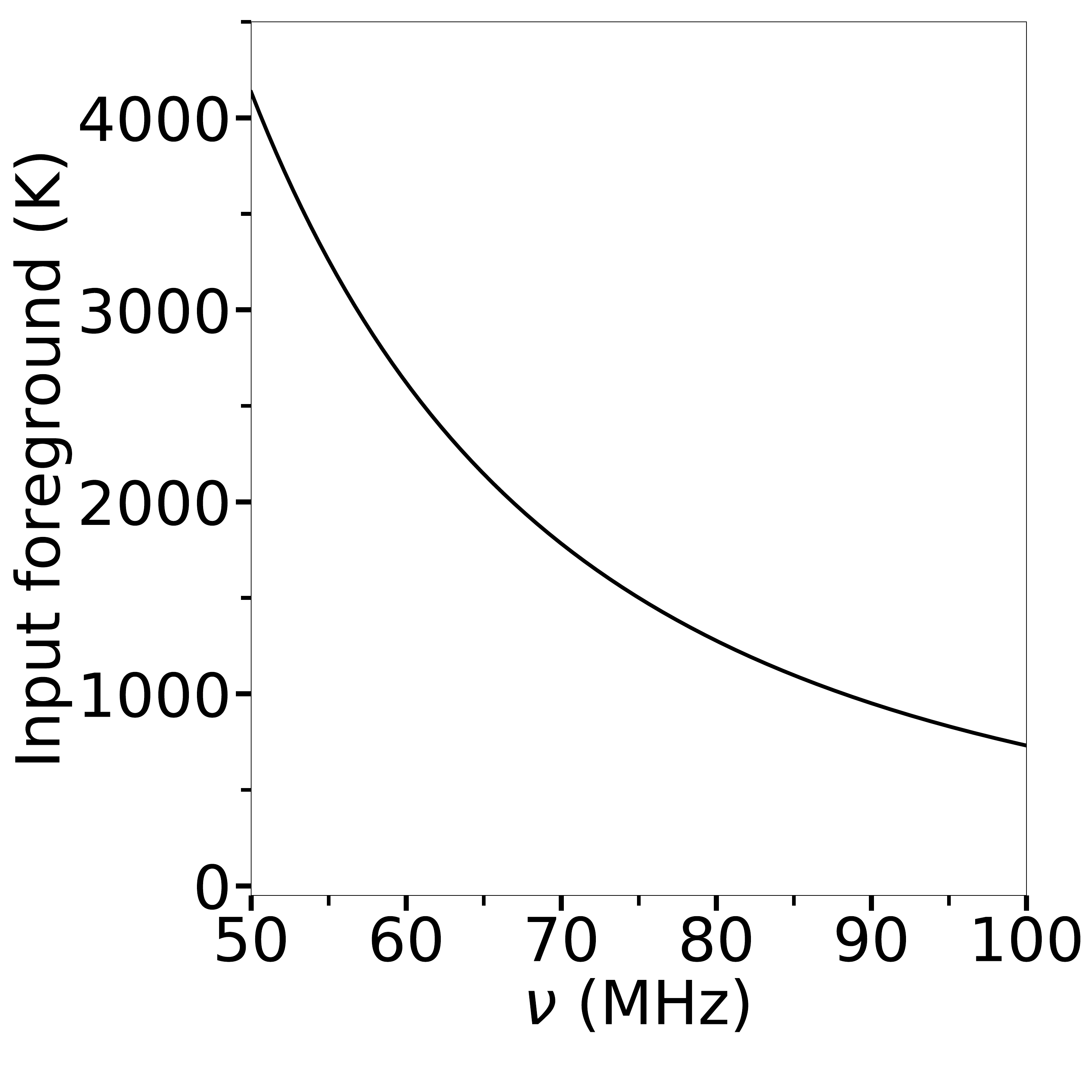}
    \includegraphics[width=0.32\textwidth]{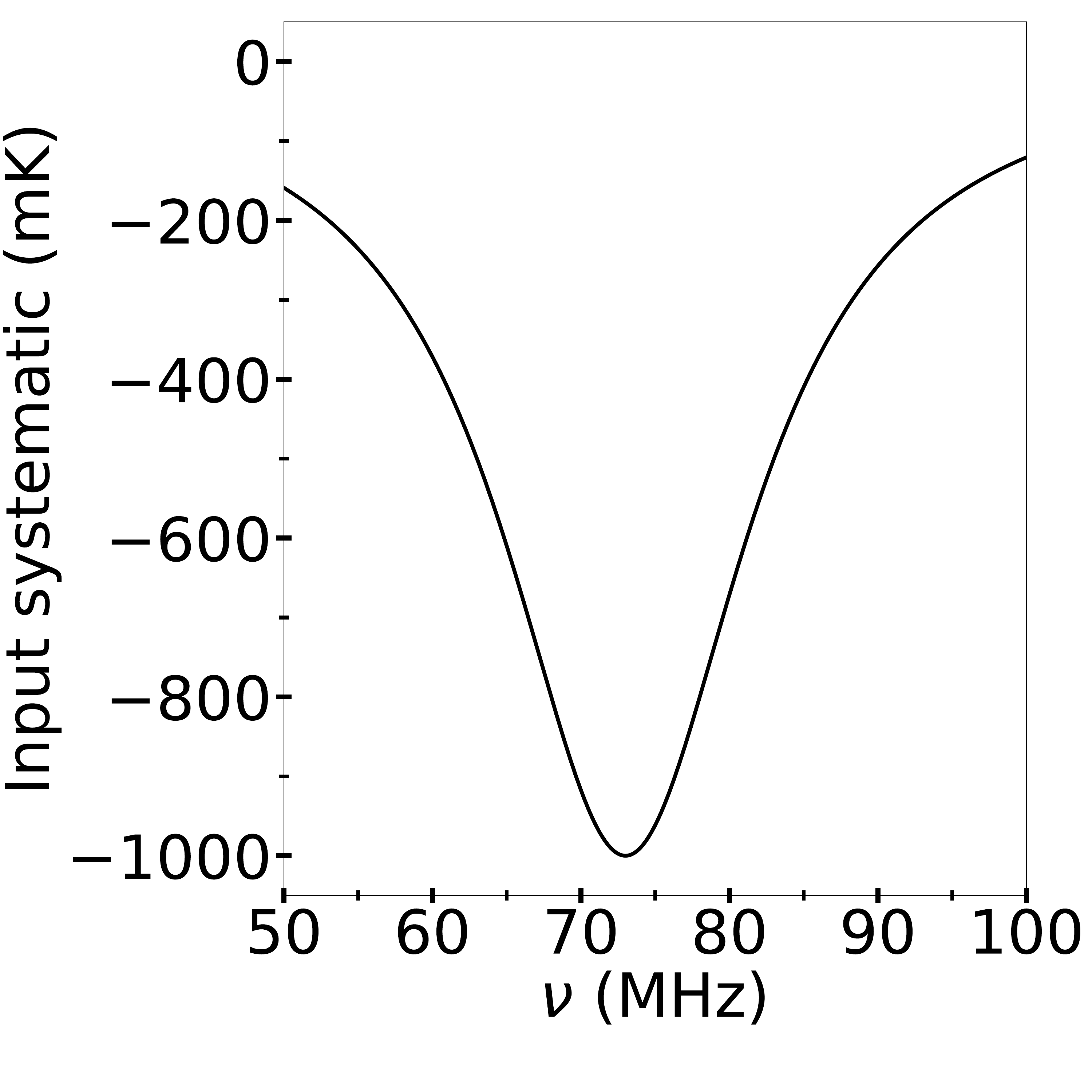}
    \includegraphics[width=0.32\textwidth]{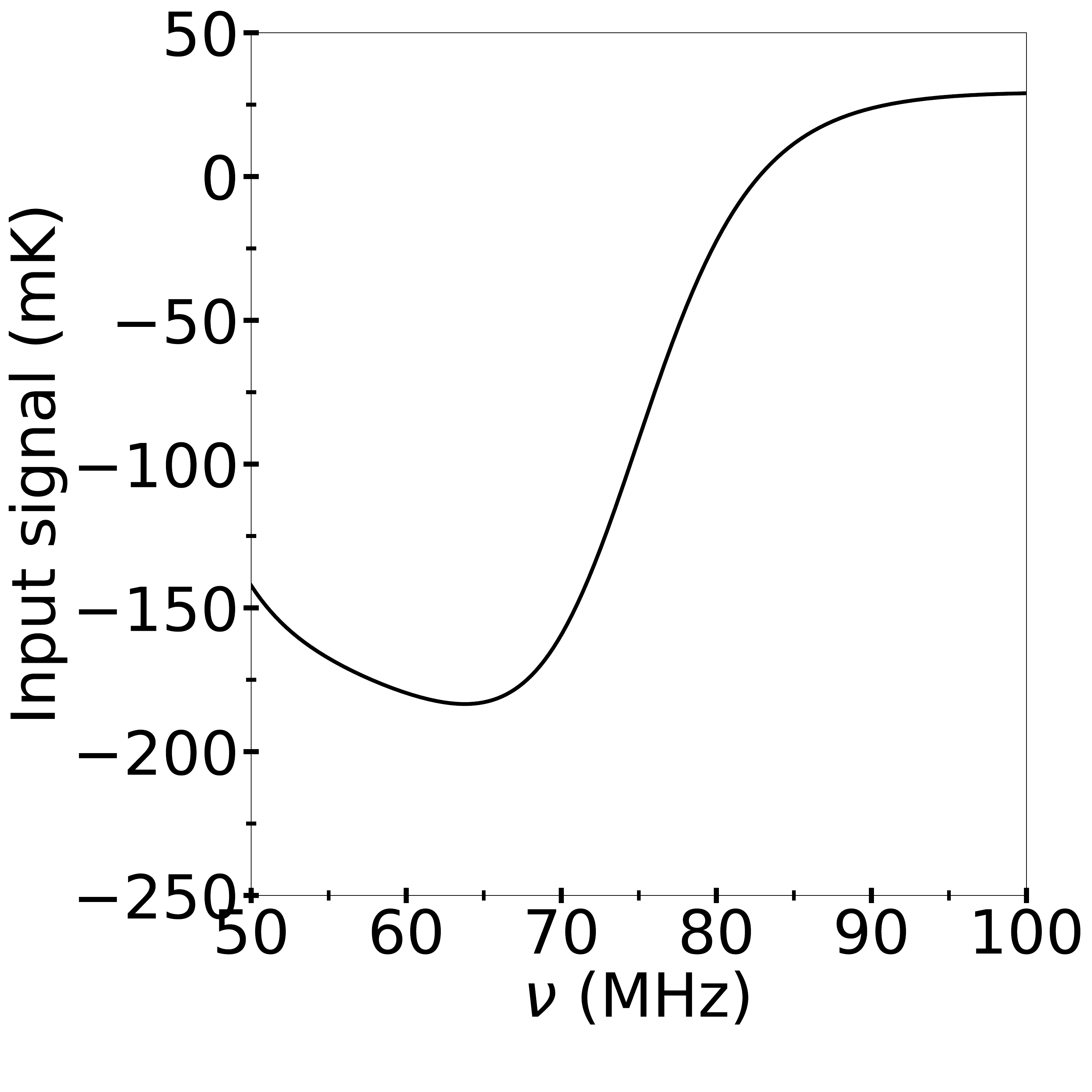}
    \caption{Input foreground, systematic, and signal used for fits performed in Section~\ref{sec:21-cm-cosmology}. The input foreground is a power law with spectral index $-2.5$, the input systematic is a Lorentzian (as defined in Equation~\ref{eq:lorentzian}) with $A=-1$ K, $\mu=73$ MHz, and $\sigma=10$ MHz (as suggested in Bradley et al., in prep.) and the input signal is a realization of the tanh model described in \cite{harker-2016}. Each of the $10^4$ simulated data vectors used for the fits is given by the sum of the foreground, systematic, signal, and Gaussian noise corresponding to 250 hr of integration and a channel width of 10 kHz.} \label{fig:input-curves}
  \end{figure}

  The input component curves to be fitted are shown in Figure~\ref{fig:input-curves}. The input foreground (left panel) is a power law with spectral index $-2.5$; so, the model in Equation~\ref{eq:power-law-times-polynomial} can fit it exactly with only one nonzero coefficient, although we use $N_{\text{terms}}=5$ for the modeling in this paper to better simulate what has been done using real data \citep{bowman-2018}. We compute the input signal (right panel) using the tanh model mentioned above. For the input systematic (center panel), we employ a Lorentzian (Equation~\ref{eq:lorentzian}) with $A=-1$ K, $\mu=73$ MHz, and $\sigma=10$ MHz, inspired by the data fitting results in Bradley et al. (in prep.).

\begin{figure}[h]
\settoheight{\tempdima}{\includegraphics[width=.31\linewidth]{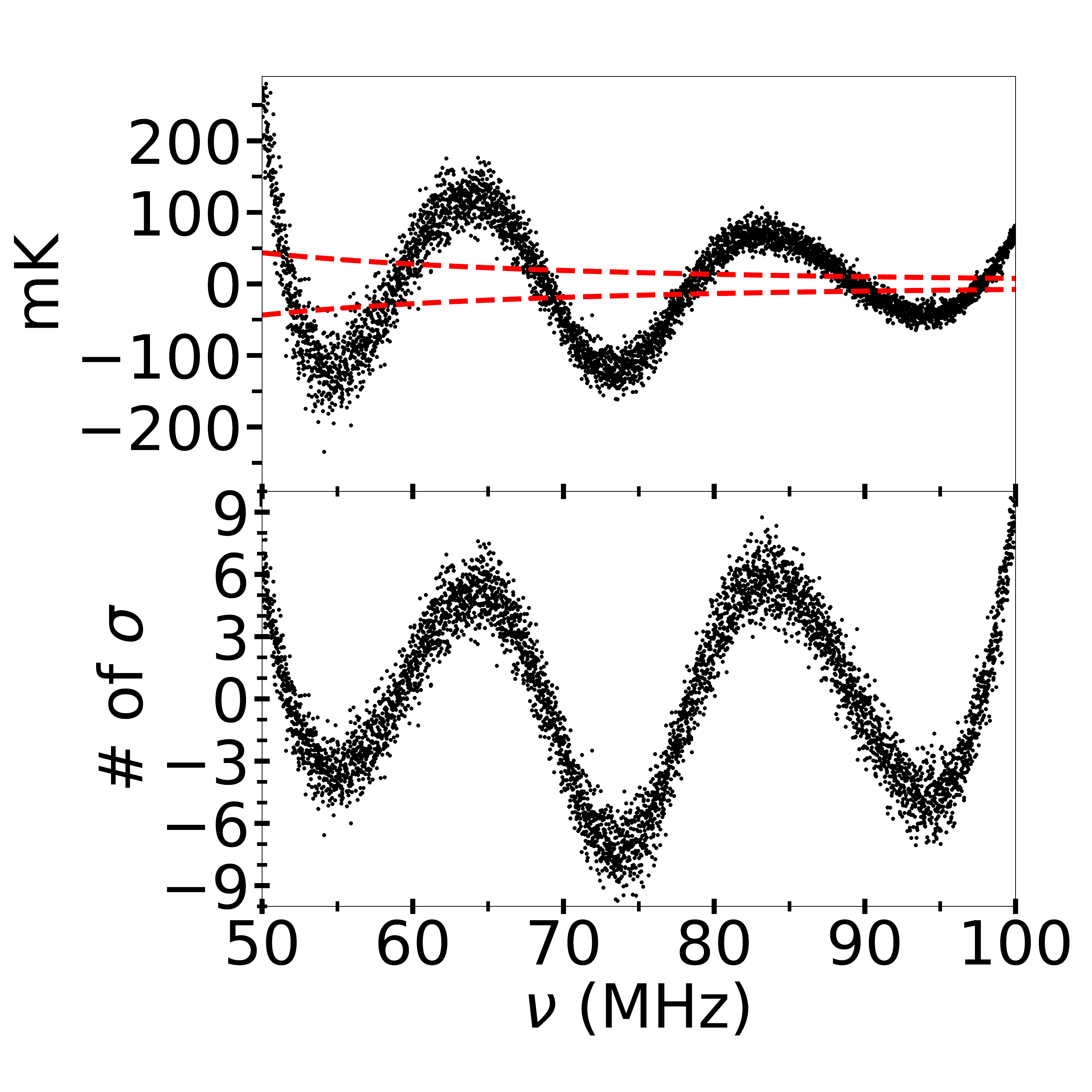}}%
\centering\begin{tabular}{@{}c@{ }c@{ }c@{ }c@{}}
&\textbf{Residuals} & \textbf{Triangle plot} & \textbf{CDF} \\
\rowname{FG only}&
\includegraphics[width=.31\linewidth]{Fig8a.pdf}&
\includegraphics[width=.3\linewidth]{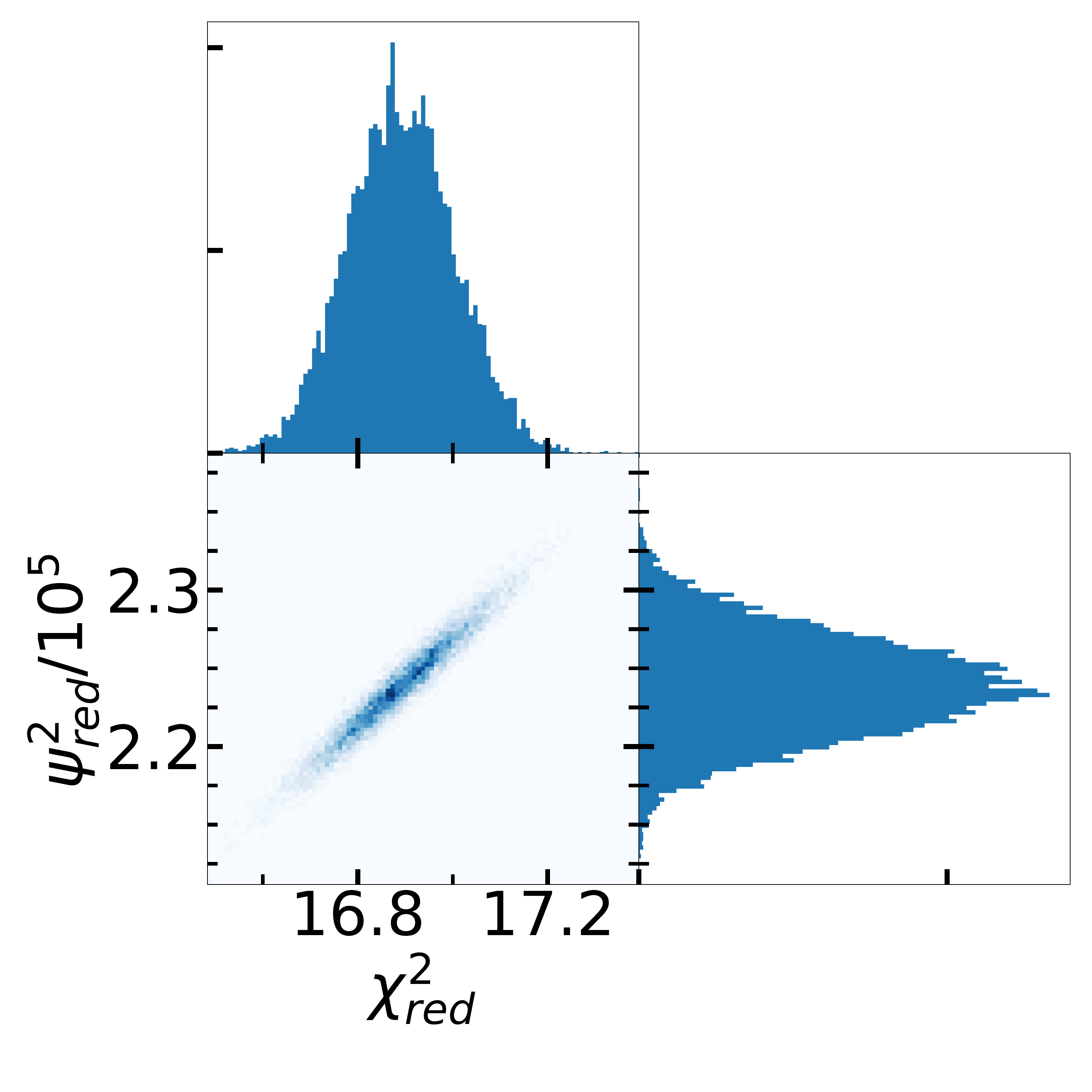}&
\includegraphics[width=.31\linewidth]{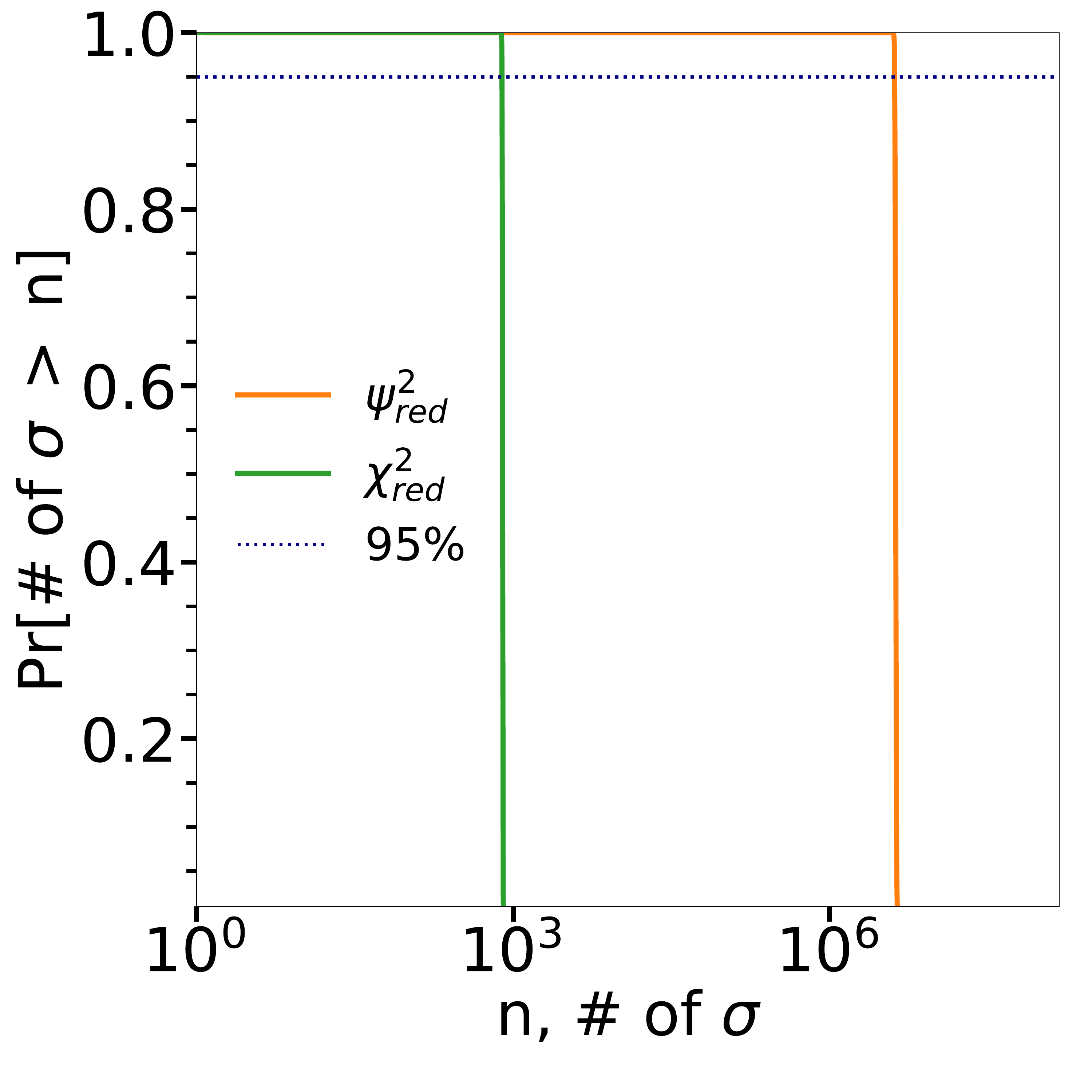}\\
\rowname{FG+Sys}&
\includegraphics[width=.31\linewidth]{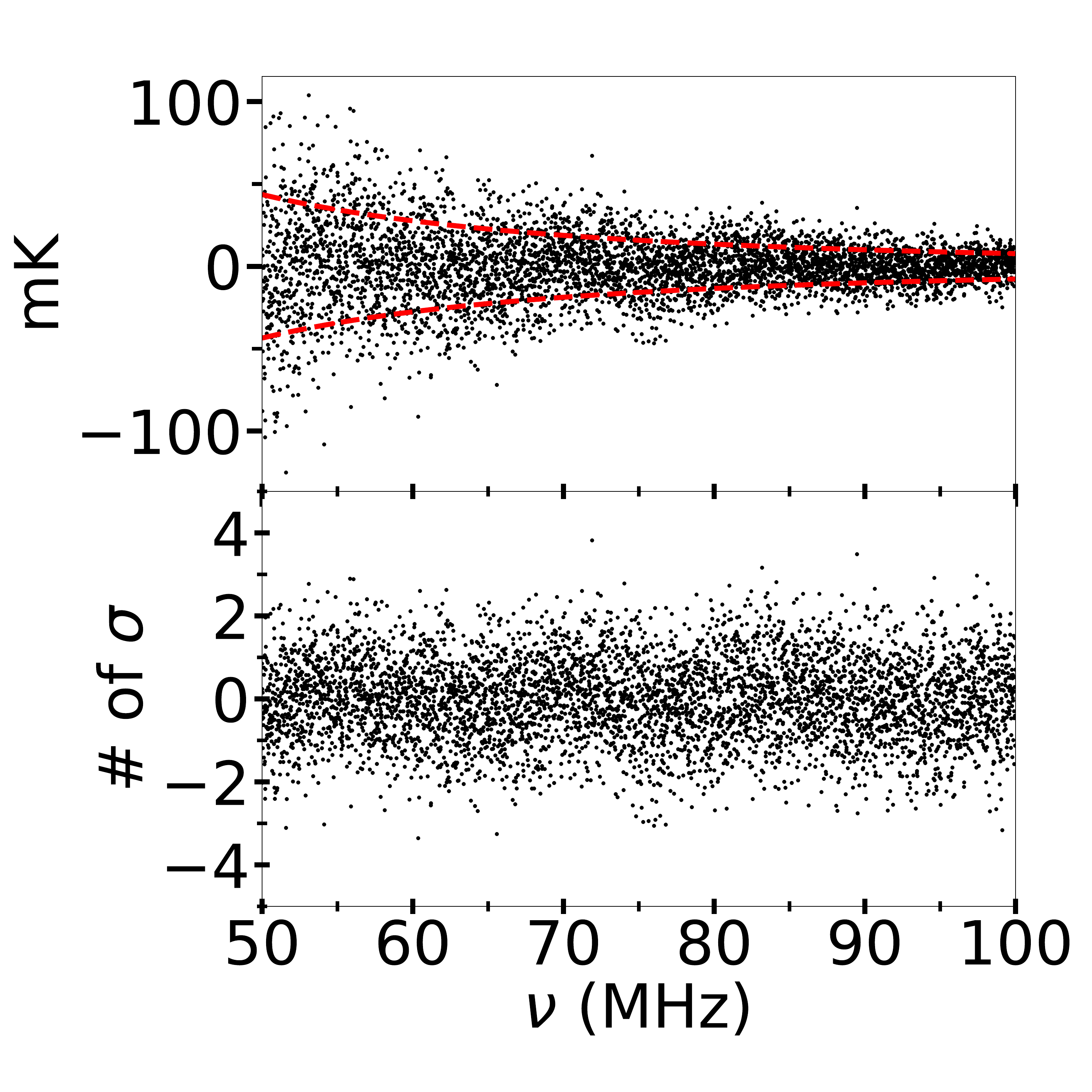}&
\includegraphics[width=.3\linewidth]{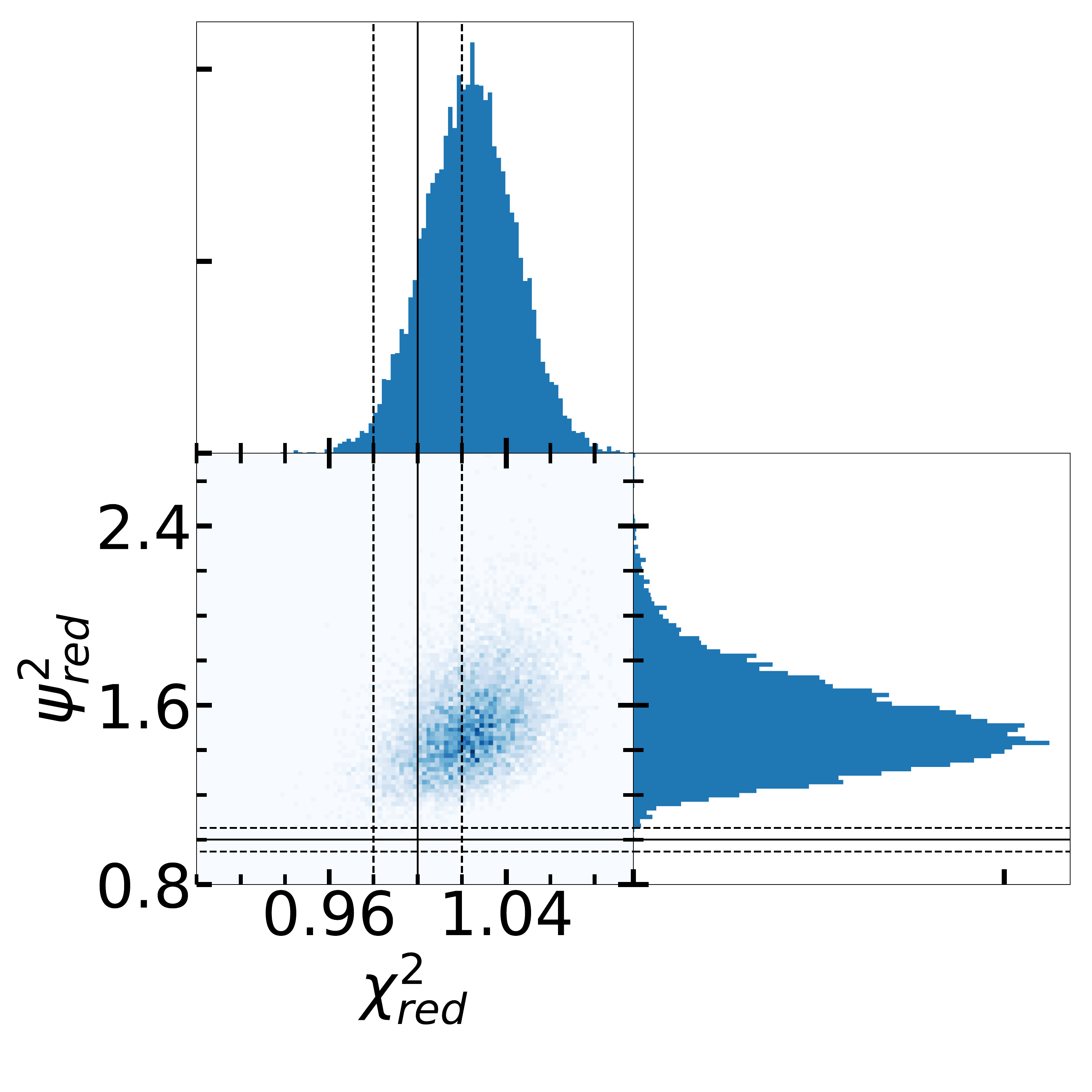}&
\includegraphics[width=.31\linewidth]{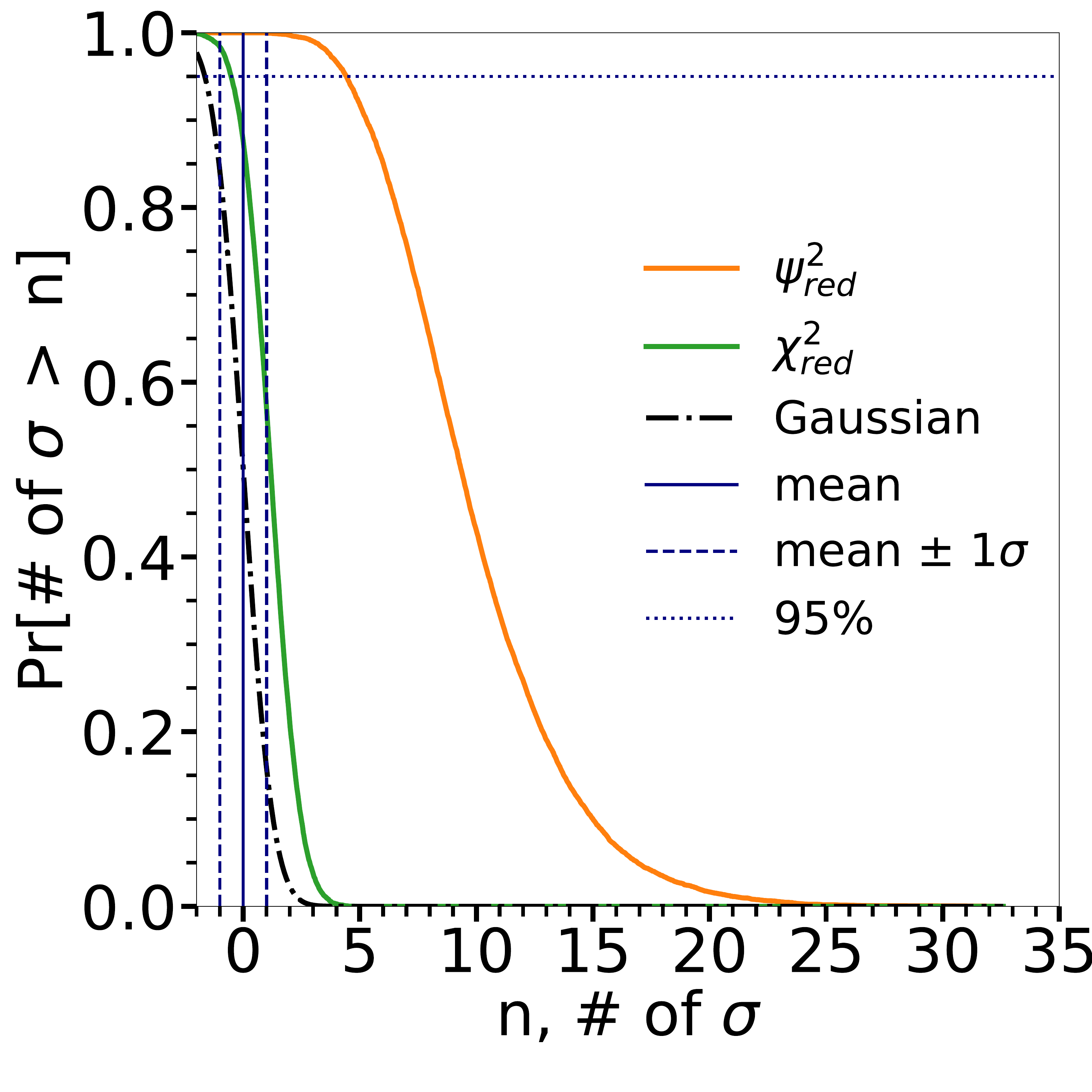}\\
\rowname{Full}&
\includegraphics[width=.31\linewidth]{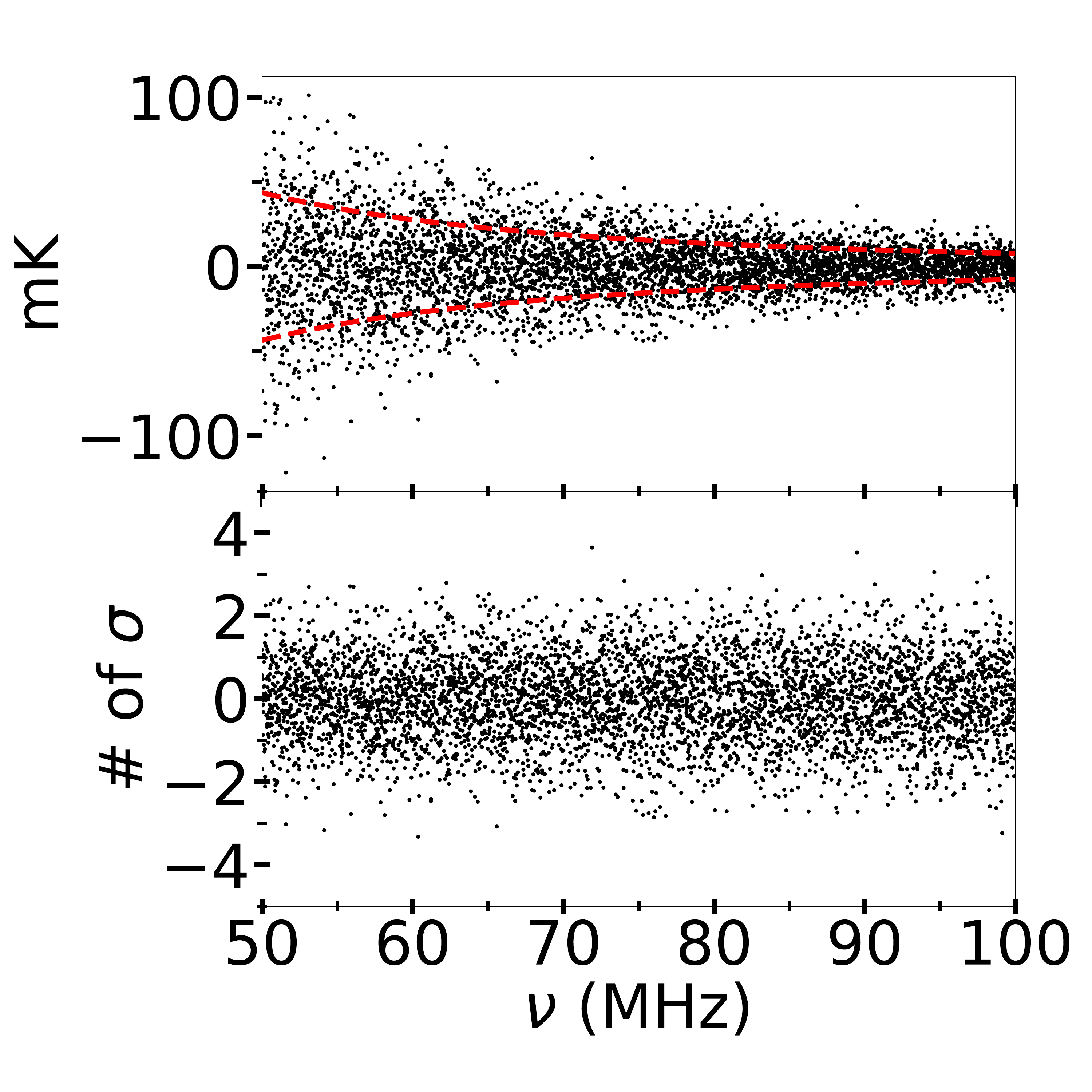}&
\includegraphics[width=.3\linewidth]{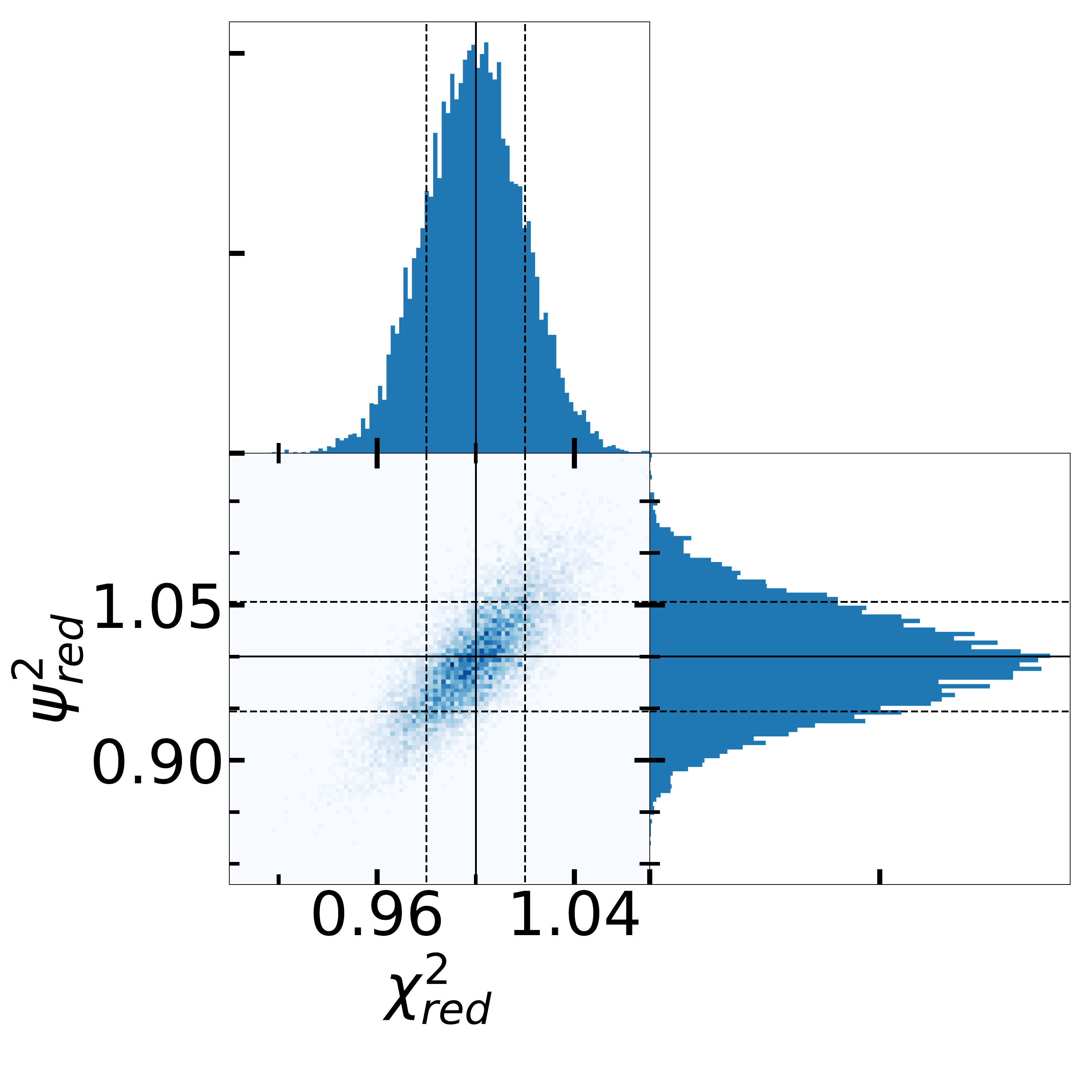}&
\includegraphics[width=.31\linewidth]{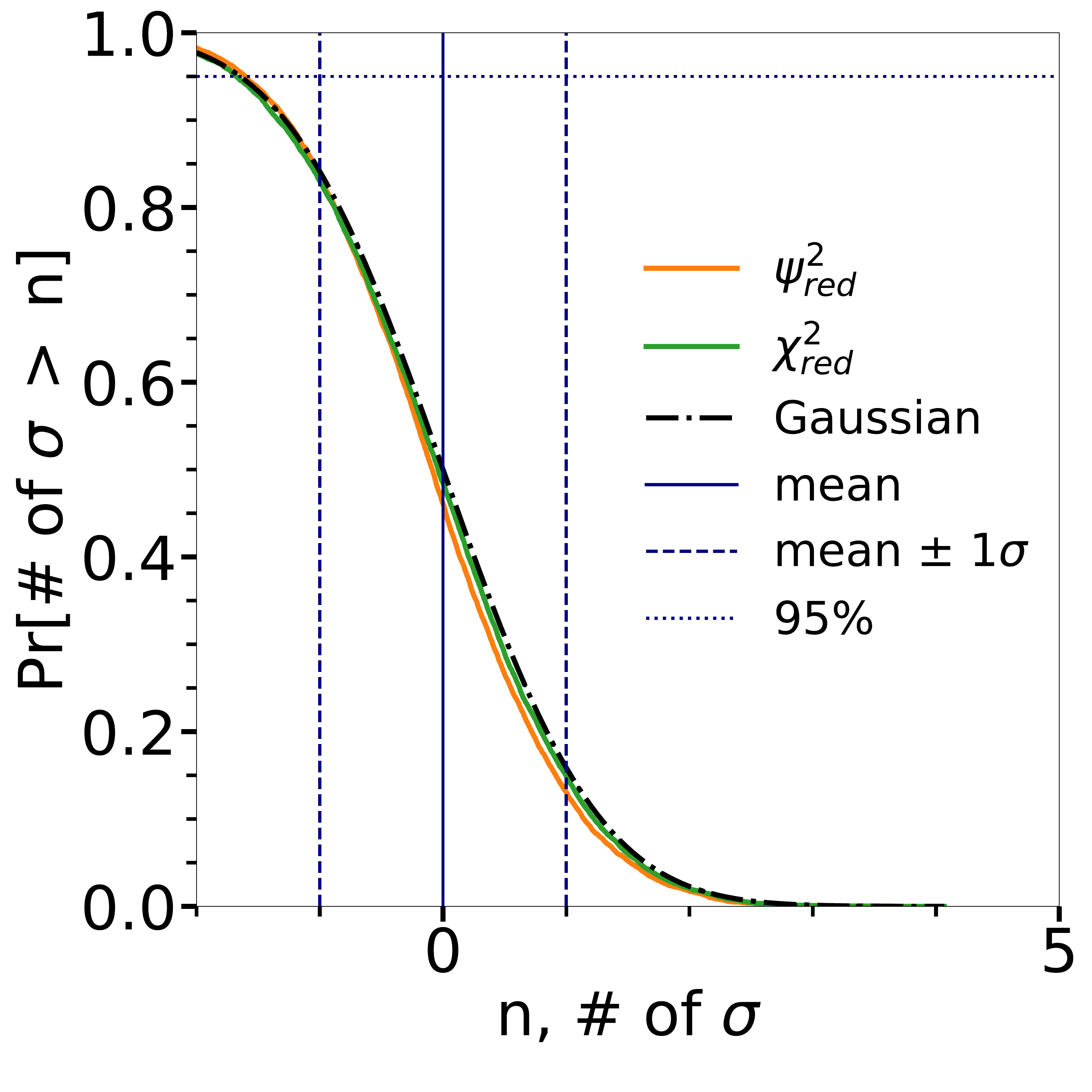}\\
\end{tabular}
  \caption{\textit{Left}: The residuals of fits to each of the data vectors described in Section~\ref{sec:21-cm-cosmology} with input components from Figure~\ref{fig:input-curves}. The dashed, red lines plotted with the residuals show the $1\sigma$ noise level. All three panels in the left column concern the same noise realization. \textit{Center}: The distributions of $\chisquaredred$ and $\psisquaredred$ for the fits. \textit{Right}: The probability that $\chisquaredred$ and $\psisquaredred$ are more than $n\sigma$ away from 1 as a function of $n$. \textit{Top}: Fit with a model given by Equation~\ref{eq:power-law-times-polynomial} with $N_{\text{terms}}=5$. The realization in the left panel has $\chisquaredred=16.9$ ($\chisquaredred-1=794\,\sigma_{\chisquaredred}$) and $\psisquaredred=224885$ ($\psisquaredred-1=4.25\times 10^6\sigma_{\psisquaredred}$). The correlation between $\psisquaredred$ and $\chisquaredred$ is 0.97. Note the log scale of the x-axis in the top right panel. \textit{Middle}: Same as top panels except that in this fit the Lorentzian systematic component is modeled simultaneously with the linear foreground model. The residual realization shown in the left panel has $\chisquaredred=1.002$ ($\chisquaredred-1=0.12\,\sigma_{\chisquaredred}$) and $\psisquaredred=1.261$ ($\psisquaredred-1=4.94\,\sigma_{\psisquaredred}$). The correlation between $\chisquaredred$ and $\psisquaredred$ is 0.40. \textit{Bottom}: Same as middle panels except that the 21-cm signal model is fitted simultaneously with the models for the linear foreground and the Lorentzian systematic. The residual realization shown in the left panel has $\chisquaredred=0.982$ ($\chisquaredred-1=-0.880\,\sigma_{\chisquaredred}$) and $\psisquaredred=0.961$ ($\psisquaredred-1=-0.733\,\sigma_{\psisquaredred}$). The correlation between $\chisquaredred$ and $\psisquaredred$ is 0.78.} \label{fig:21-cm-application-fits}
\end{figure}

\subsection{Foreground-only fit} \label{sec:foreground-only-fit}
  
  The $1^{\text{st}}$ row of Figure~\ref{fig:21-cm-application-fits} shows the statistical properties of fits performed with only the foreground model of Equation~\ref{eq:power-law-times-polynomial} with $N_{\text{terms}}=5$. Since the model in this case is completely linear, the fits can be performed analytically. We utilize \texttt{pylinex} to perform this task on all data realizations simultaneously. For this fit, it is clear that both $\psisquaredred$ and $\chisquaredred$ can effectively discern that the data is far from being fully modeled.
  
  \subsection{Foreground+systematic fit}
  
  It is conceivable that one would fit for only the models representing the foreground emission and the instrumental systematic feature, unsuspectedly confusing the systematic for a feature like the 21-cm signal in the data. Results of such fits which leave out the signal model are shown in the $2^{\text{nd}}$ row of Figure~\ref{fig:21-cm-application-fits}. These panels illustrate a key outcome of this work. While the $\chisquaredred$ values are somewhat skewed above 1, most realizations are close enough to 1 for a $\chisquaredred$ test to be unable to reject with high confidence the NH that the residuals are purely noise. On the other hand, the values of $\psisquaredred$ are much farther from 1 in units of $\sigma$, indicating that, in this situation, $\psisquaredred$ can be used to conclude with high confidence that there are unmodeled, non-random components remaining in the data. Such an ability should prove $\psisquaredred$ to be a powerful new statistical tool for analyzing ongoing and upcoming 21-cm data sets.
  
  \subsection{Full fit}

  The bottom panels of Figure~\ref{fig:21-cm-application-fits} show the results corresponding to a fit where the model accounts for all three non-random data components---foreground, systematic, and signal---and they are all fitted simultaneously. In this case, since the data are sufficiently represented by the model, $\psisquaredred$ and $\chisquaredred$ have distributions very similar to those they have under the NH, where the residual is pure noise (compare the bottom panels of Figure~\ref{fig:21-cm-application-fits} to those of Figure~\ref{fig:null-hypothesis}). Usefully, this confirms that the goodness-of-fit hypothesis test proposed in Section~\ref{sec:hypothesis-test} is reasonable, since it relies on the orange solid and gray dash-dot curves in the bottom right panel of Figure~\ref{fig:21-cm-application-fits} being essentially identical.

\section{Discussion and conclusions} \label{sec:discussion}

  This paper has laid out the theory behind a new goodness-of-fit statistic, $\psisquaredred$. For a large number of data points and residuals that are noise-like, as they are after a good fit, $\psisquaredred$ has a simple distribution similar to the distribution of $\chisquaredred$, making it as simple to use for testing whether a data vector has been modeled down to the noise floor. However, $\psisquaredred$ is much more sensitive to low-level, wide-band residual features than $\chisquaredred$. Since plenty of astrophysical observations yield data that have smooth systematics spanning the entirety of the data (e.g. foreground emission in radio astronomy), $\psisquaredred$ has multiple potential applications across the field where it can provide significant additional discriminating power for determining the presence of unfitted residual structures in the data (as demonstrated in the middle panels of Figure~\ref{fig:21-cm-application-fits}).

  The following are possible topics for future work in developing the theory of the $\psisquaredred$ statistic:
  \begin{itemize}
    \item
      \emph{Exact, closed-form expression of the variance of $\psisquaredred$ under the NH}: While currently unknown, it might be possible to obtain using Equation~\ref{eq:variance}.
    \item
      \emph{Effects and corrections for non-negligible $\frac{p}{N}$}: In this paper, we have dealt in the asymptotic limit where the number of channels $N$ is much greater than the number of parameters $p$. But, the behavior of $\psisquaredred$ away from this limit is unknown.
    \item
      \emph{Non-Gaussianity}: Throughout this paper, we assumed that the noise in the data which generated the residuals under test had a Gaussian distribution. While this assumption is somewhat restrictive, it can be relaxed if the corresponding results are recalculated. A list of the assumptions made for key results in this paper is given in Appendix~\ref{app:gaussianity-assumptions}. From these assumptions, it is clear that for instance in the case of experiments whose data are Poisson distributed, to name another common example in astrophysics, the NH distribution of $\chisquaredred$ and $\psisquaredred$ must be recalculated before using them as goodness-of-fit tests. This would allow a broader application of these statistics.
    \item
      \emph{Covariance estimation}: The definitions in Section~\ref{sec:methods} rely on a known covariance matrix $\bC$. However, in many real-world data applications, the covariance matrix of the data is unclear a priori, and must be estimated from the data themselves. The effects on $\psisquaredred$ from estimating $\bC$ while simultaneously fitting the data remain unexplored.
    \item
      \emph{Correlation normalization effects}: Imprecision in the noise level used in normalizing the residuals can have a significant impact on $\psisquaredred$ (see Appendix~\ref{app:noise-level-imprecision}), mostly because it is a quartic, not quadratic, function of the residuals. One way to avoid this reliance on a precise knowledge of the noise level is to change the definition of the correlations, $\rho_q$, by dividing them by $\chisquaredred$. This is equivalent to normalizing $\psisquaredred$ by dividing it by $(\chisquaredred)^2$. It is not yet known how this form of normalization affects the distribution of $\psisquaredred$, although preliminary simulations indicate that when residuals are pure white noise, $\psisquaredred/(\chisquaredred)^2$ has a variance less than $\frac{14}{N}$ and is, remarkably, nearly uncorrelated with $\chisquaredred$.
    \item
      \emph{Binning effects}: Binning is a common analysis technique used to reduce the number of data channels, sometimes applied for purely computational reasons. In any analysis technique, though, one should require that the sensitivity to residual features remains the same. In the case of $\chisquaredred$, this can be achieved with binned data if the residual features are well resolved after binning and the original data points were evenly spaced. It could be, however, that the sensitivity of $\psisquaredred$ to correlations is changed after binning even in cases where $\chisquaredred$ remains unaffected. In addition, there is no rigorous, straightforward method of binning data which can guarantee in general that no relevant information is lost. Indeed, with basic binning methods, this can only be ensured when the non-random component is linear across channels. A binning procedure that leaves the informational content of correlations quasi-invariant would allow combining this popular technique with $\psisquaredred$.
    \item
      \emph{Instrumental effects}: Instruments which measure real data are imperfect and can affect $\psisquaredred$ through three main mechanisms:
      \begin{itemize}
          \item
            \emph{Smooth systematic effects}: Systematics introduce the need for additional models which must be fit alongside other aspects of the data. The impact of some of these types of features on $\psisquaredred$ was the subject of Section~\ref{sec:fitting}. But, other 21-cm cosmology systematics and residual features common in other fields must be studied individually to determine the sensitivities of $\psisquaredred$ and $\chisquaredred$ in each fitting case.
        \item
          \emph{Correlations among nearby channels}: One plausible way of handling short-spacing correlations induced by the instrument is by imposing a minimum value of $q$ in the sum defining $\psisquaredred$ in Equation~\ref{eq:psisquared-definition}. A maximum value of $q$ may also be desired to remove the effect of large-scale correlations for which, by construction, there are few samples. The modifications on $\psisquaredred$ caused by imposing such bounds on the spacings it includes is unknown.
        \item
          \emph{Corrupted, defective, or saturated channels}: Whether due to hardware malfunction or natural unwanted oversaturation (e.g. RFI in radio astronomy), some channels need to be removed from the data before analysis, leading to missing data. While missing data does not change $\chisquaredred$ because it only depends on the magnitudes of each individual residual point, it could significantly impact $\psisquaredred$.
      \end{itemize}
  \end{itemize}
  
  In 21-cm cosmology, and potentially in other applications, it is expected that all individual spectra contain the same desired spectral signal while features due to systematics can vary. One possible way to test if a wide-band, cross-spectra signal exists in the data is to examine the frequency space correlations $\rho_q$ averaged over spectra and compute a $\psisquaredred$ statistic corresponding to these averages. Alternatively, one could examine the frequency correlations and $\psisquaredred$ statistic for an averaged spectrum. In both cases, the averaging over spectra should downweight features which exist in only one spectrum while reinforcing features which exist in all spectra. Such analyses promise to be of high value to identify low amplitude-to-noise signals in the presence of similarly wide-band systematics and/or even other signals. 
  
  Utilizing $\psisquaredred$, together with $\chisquaredred$, will therefore provide a deeper view of the theoretical ability to describe measurements in any research field evaluating the performance of model fits to data.

\section*{Appendices}
\appendix

\section{Practical computation of $\psisquaredred$} \label{app:practical-computation}
  
  Here, we summarize a prescription for calculating $\psisquaredred$ from a given data vector $\by$ and a corresponding noise covariance estimate $\bC$:
  \begin{enumerate}
  \item Perform a fitting procedure of your choosing to obtain a residual $\bdelta$ from $\by$ and $\bC$.
  \item Form the normalized residual, $\bDelta$, via $\bDelta=\bC^{-1/2}\bdelta$. If $\bC$ is diagonal, use $\Delta_k=\frac{\delta_k}{\sigma_k}$.
  \item Compute the correlations $\rho_q$ for nonzero $q$ via Equation~\ref{eq:correlation-definition}.
  \item Normalize the correlations $\rho_q$ by dividing by their noise levels, $\sqrt{\frac{1}{N-q}}$.
  \item $\psisquaredred$ is the mean square of these normalized correlations.
  \end{enumerate}
  The last two steps are encapsulated in Equation~\ref{eq:psisquared-definition}. In Python, calculation of the $\psisquaredred$ statistic is implemented in \texttt{psipy},\footnote{\url{https://bitbucket.org/ktausch/psipy}} a script which defines functions for $\psisquaredred$ (with some modifications suggested in Section~\ref{sec:discussion} included as options) and $\chisquaredred$. The exact same functions are built into the larger \texttt{pylinex}\footnote{See Footnote \ref{footnote:pylinex-link}.} code, which was used to perform all of the fitting for this paper.
  
\section{$\psisquaredred$ sensitivity to spikes} \label{app:spike-sensitivity}

  As described in Footnote~\ref{footnote:stronger-inequality}, it can be derived from Equation~\ref{eq:full-psisquaredred} that when the noise covariance is accurate, the expectation value of $\psisquaredred$ satisfies the inequality 
  \begin{equation}
      \E[\psisquaredred]\ge 1+\overline{\psisquaredred}+\frac{\bmu^T\bW\bmu}{N-1}, \label{eq:app-inequality}
  \end{equation}
  where $\bmu$ is the non-random component in the vector being tested with a noiseless $\psisquaredred$ given by $\overline{\psisquaredred}$, and $\bW$ is a diagonal matrix whose nonzero elements have the form
  \begin{equation}
    W_{nn}=H_{\lceil(N+n)/2\rceil-1}+H_{N-\lfloor(n+1)/2\rfloor}-H_{n-1}-H_{N-n}.
  \end{equation}
  
  \begin{figure}[t!!]
    \centering
    \includegraphics[width=0.32\textwidth]{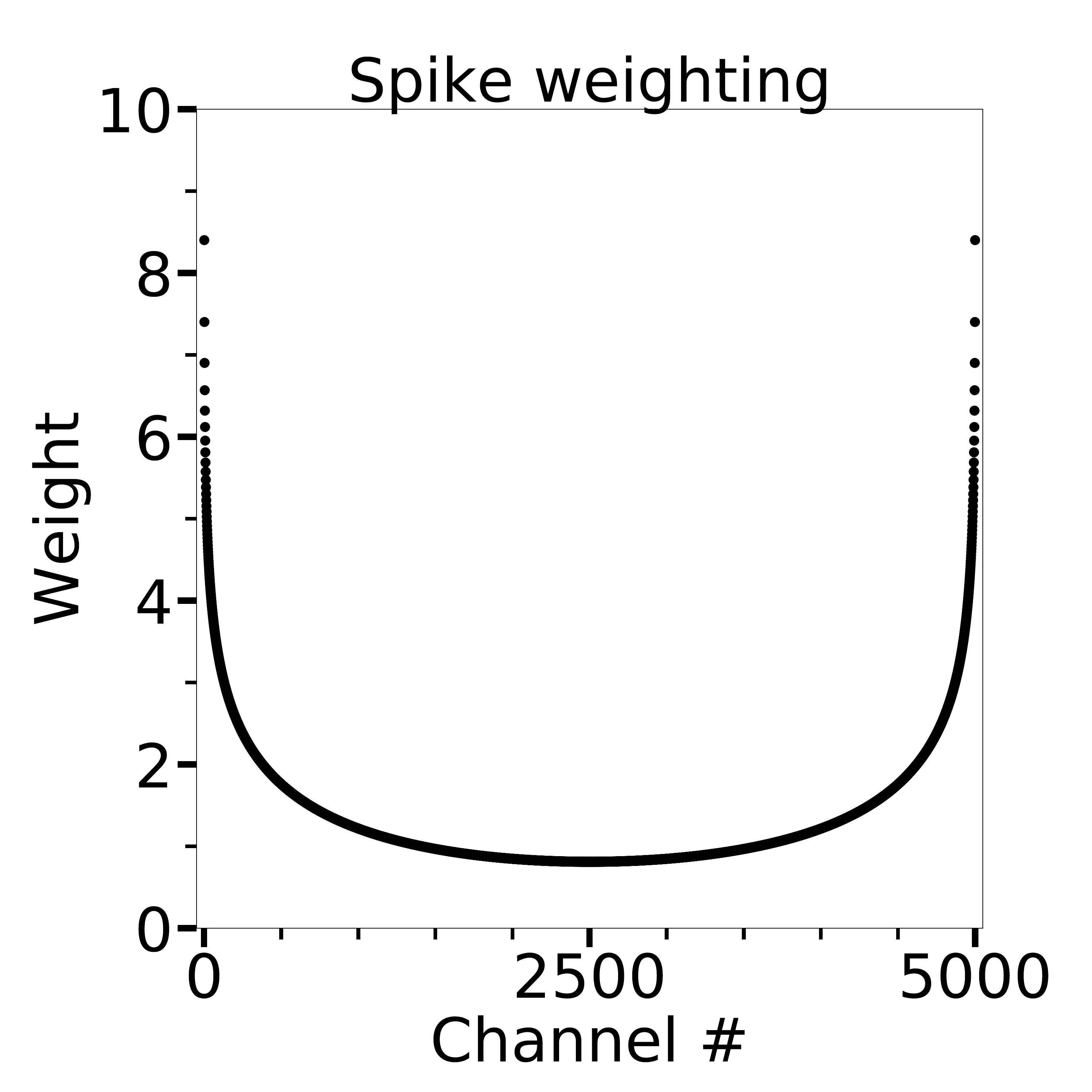}
    \includegraphics[width=0.32\textwidth]{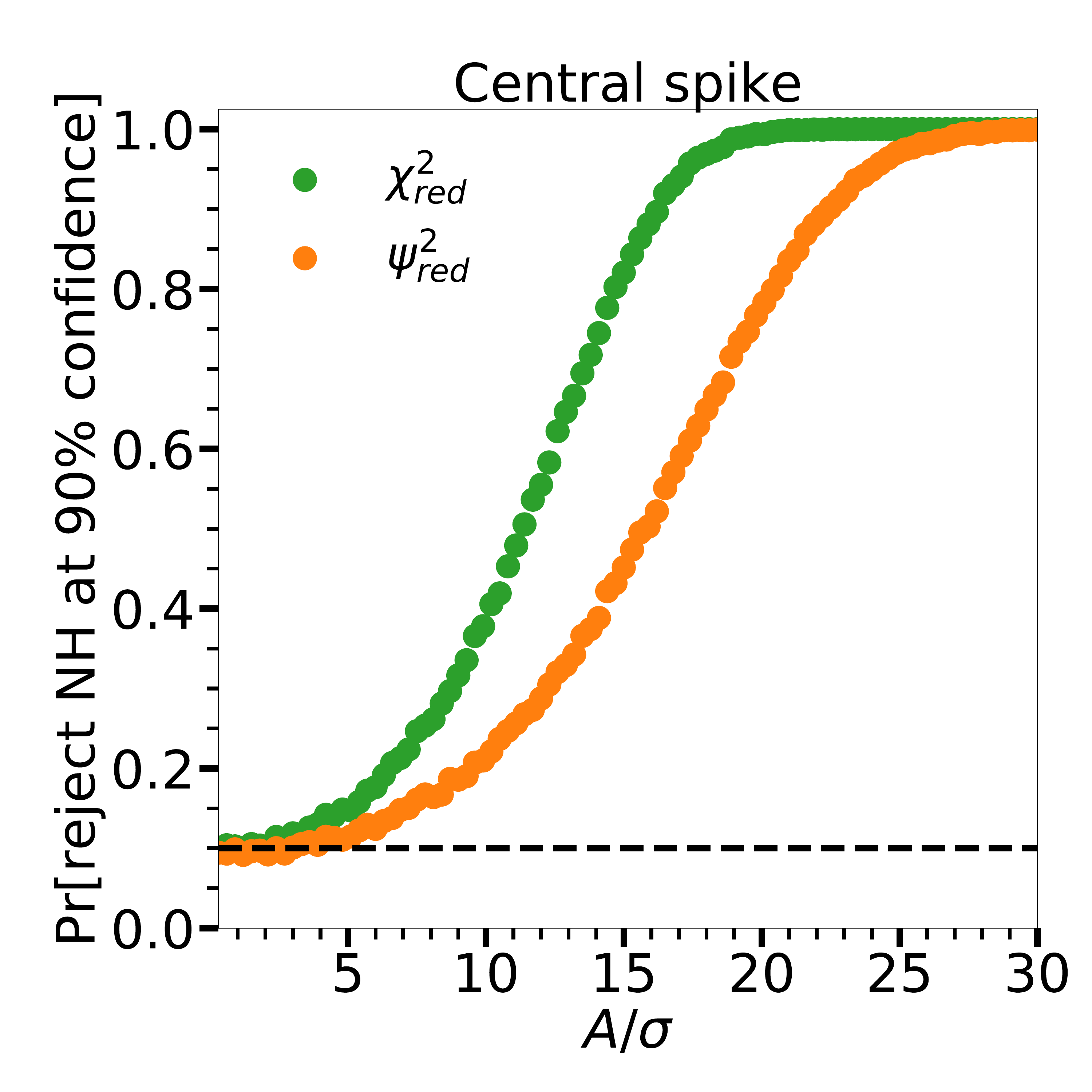}
    \includegraphics[width=0.32\textwidth]{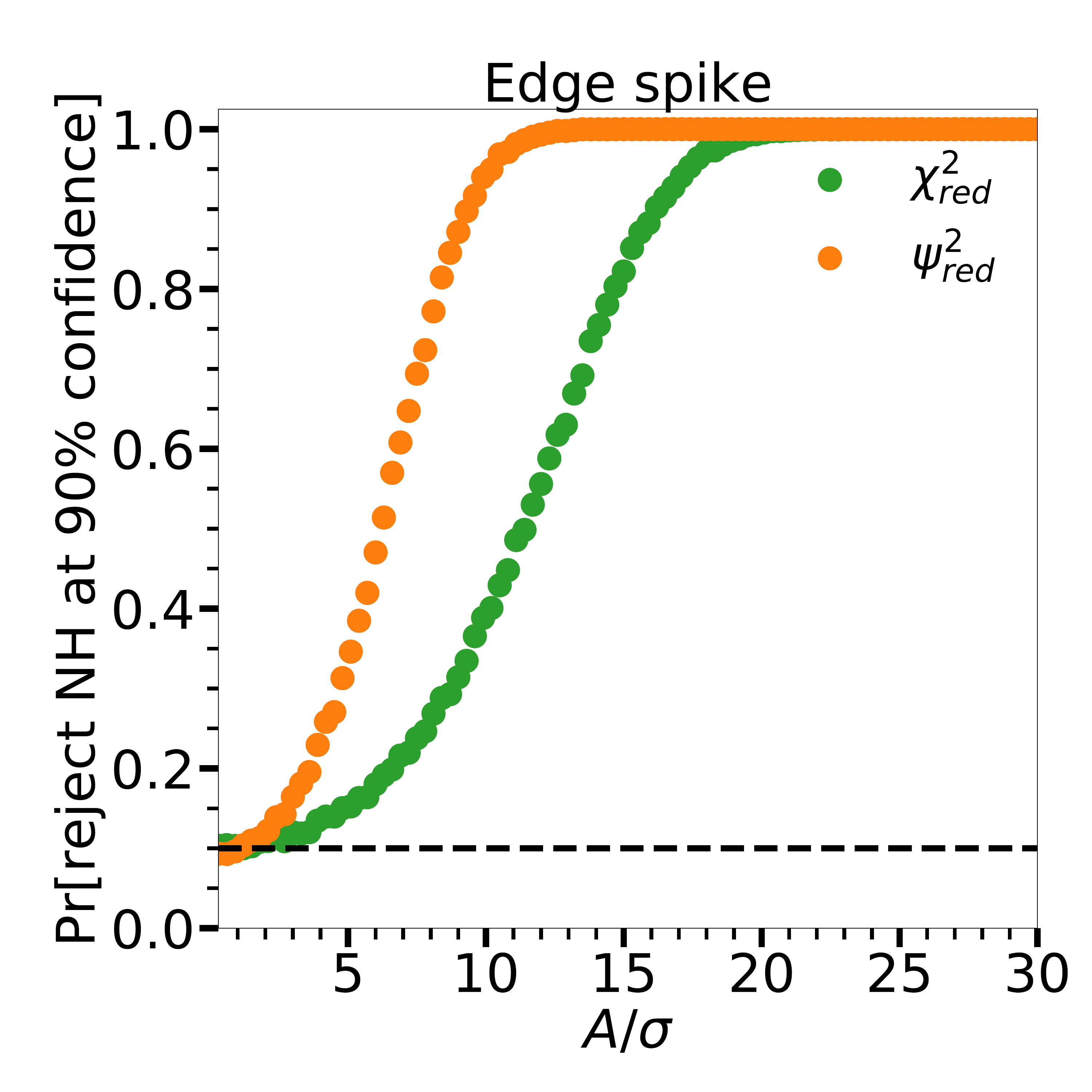}
    \caption{\textit{Left}: Weights on non-noise spikes as a function of channel number when calculating the mean of $\psisquaredred$. The y-axis is normalized by the weight put on the channels by $\chisquaredred$. The weights assigned to channels in the middle of the band are roughly $\varphi=\ln{\left(\frac{9}{4}\right)}\approx 0.811$ while the weights assigned to the channels at the edges are approximately $(\ln{N}+\gamma-\ln{2})$ where $\gamma$ is the Euler-Mascheroni constant. The next two panels are similar to those of Figure~\ref{fig:rejection-probability-vs-noise-level}. \textit{Center}: Probability to reject NH at $90\%$ confidence when in the center of the band a non-random spike of amplitude $A$ is added to noise of magnitude $\sigma$. As with Figure~\ref{fig:rejection-probability-vs-noise-level}, the probabilities shown arise from simulations of $10^4$ noise realizations with $5001$ data points at $100$ different noise levels. \textit{Right}: Same as center except that the spike which was used as the non-random component was in the first channel instead of the central channel. $\psisquaredred$ is more sensitive to spikes at the edges of the band than $\chisquaredred$ but less sensitive than $\chisquaredred$ at detecting spikes in the center of the band (see the center panel).} \label{fig:spike-weights}
  \end{figure}
  
  \noindent The diagonal of $\bW$ contains weights assigned to spikes at different points in the band. The left panel of Figure~\ref{fig:spike-weights} shows these weights when $N=5001$. The other two panels show the probability to reject the null hypothesis for two spikes placed at different positions of the band. We note that the weight applies only if there is exactly one spike in the band, when $\overline{\psisquaredred}=0$. When more than one channel is affected, non-random correlations are induced, causing $\overline{\psisquaredred}$ to be nonzero, so that $E[\psisquaredred]$ is higher than the weights alone would imply.
  
\section{Noise level dependence of the $\psisquaredred$-$\chisquaredred$ correlation} \label{app:correlation-noise-dependence}

\begin{figure}[t!!]
    \centering
    \includegraphics[width=0.6\textwidth]{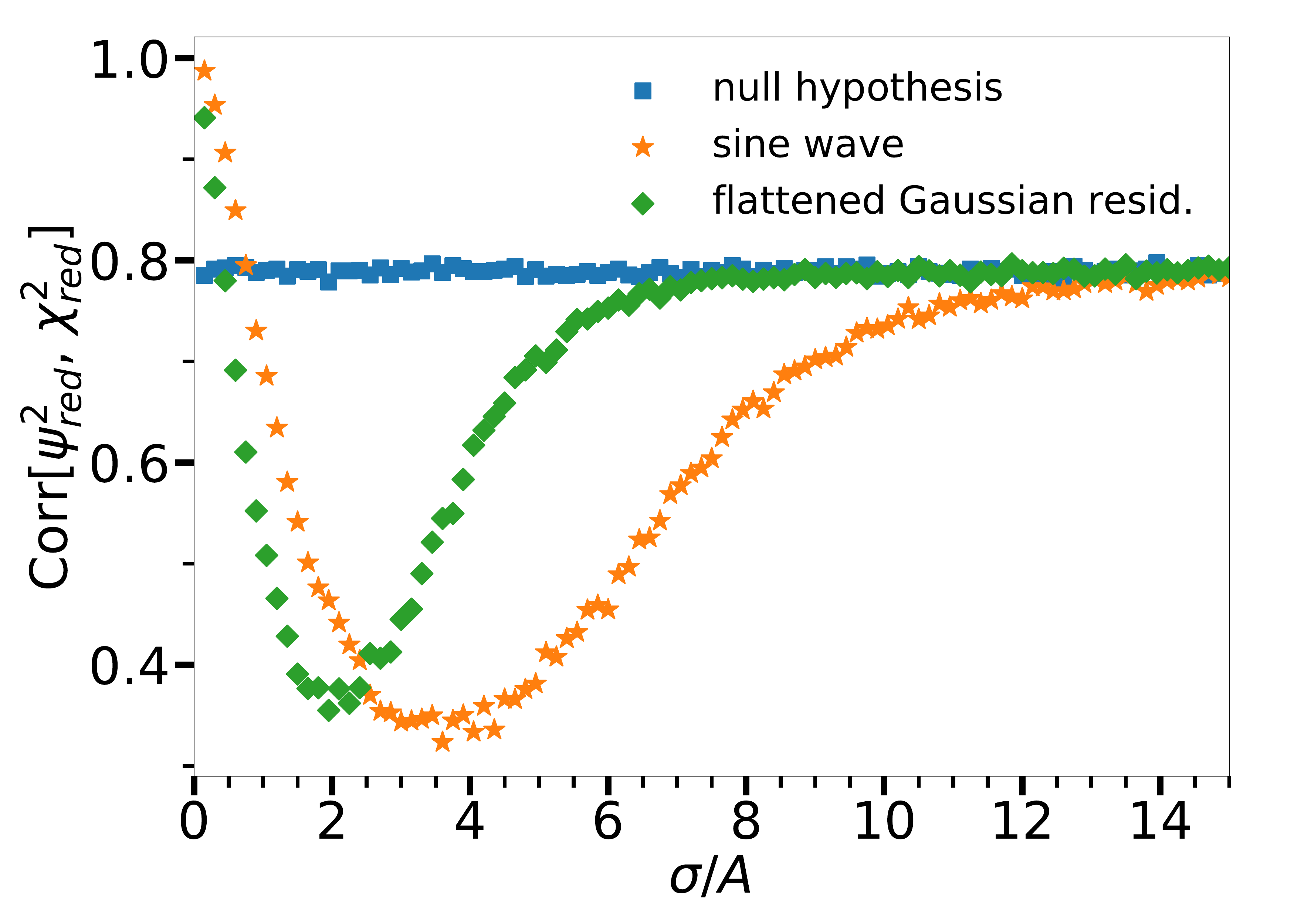}
    \caption{Correlation coefficients between $\psisquaredred$ and $\chisquaredred$ for various features as a function of noise level computed with the same samples of $\psisquaredred$ and $\chisquaredred$ as Figure~\ref{fig:rejection-probability-vs-noise-level}. The labels `null hypothesis', `sine wave', and `flattened Gaussian resid.' refer to the features from Figures~\ref{fig:null-hypothesis},~\ref{fig:nonrandom-sine-wave},~and~\ref{fig:nonrandom-ideal-residual}, respectively. Note that the values of $\frac{\sigma}{A}$ corresponding to the minimum correlations in the green and orange curves roughly match the points in Figure~\ref{fig:rejection-probability-vs-noise-level} where the NH-rejection probabilities of $\psisquaredred$ and $\chisquaredred$ start diverging when increasing $\frac{\sigma}{A}$.} \label{fig:correlation-vs-noise-level}
  \end{figure}

  The correlation between $\chisquaredred$ and $\psisquaredred$ calculated from a given data vector is a function of the  non-random feature in the vector and the noise level. Figure~\ref{fig:correlation-vs-noise-level} shows the correlation of $\chisquaredred$ and $\psisquaredred$ for the features of Figures~\ref{fig:null-hypothesis},~\ref{fig:nonrandom-sine-wave},~and~\ref{fig:nonrandom-ideal-residual} as a function of the noise level. As mentioned in Section~\ref{sec:null-hypothesis}, when NH is true (i.e. when residuals are purely noise), the correlation between $\psisquaredred$ and $\chisquaredred$ is about $0.8$ no matter the noise level. Even when there is a feature in the data, if the noise is large enough, this correlation is about $0.8$, as can be seen by noting the convergence of the points at the right side of the plot. Interestingly, for wide-band features we have examined
  \begin{equation}
  \lim_{\frac{\sigma}{A}\rightarrow 0}\Corr[\psisquaredred,\chisquaredred]=1,
  \end{equation}
  indicating that $\chisquaredred$ and $\psisquaredred$ detect unambiguous wide-band features in the same way. For such features, the $\frac{\sigma}{A}$ space between this small noise limit and the large noise limit of NH generally includes a minimum where
  \begin{equation}
    \Corr[\psisquaredred,\chisquaredred]_{\text{min}}\approx 0.35. 
  \end{equation}
  The noise level at which this correlation occurs is around the point that the NH-rejection probabilities of $\psisquaredred$ and $\chisquaredred$ begin to diverge in Figure~\ref{fig:rejection-probability-vs-noise-level} in the increasing direction of the x-axis. Note that these levels are not expected to be exactly equal because the position of the divergence of the rejection probabilities of $\psisquaredred$ and $\chisquaredred$ depends on the confidence level, whereas the noise level which produces a minimum correlation between $\psisquaredred$ and $\chisquaredred$ is independent of the confidence level.
  
\section{Effect of incorrect noise level on $\rho_q$ and $\psisquaredred$} \label{app:noise-level-imprecision}

  We now consider the case where there is no non-random component in the residuals $\bDelta$ but the noise does not have the expected profile. We denote the error level used to normalize $\delta_k$ by $(\sigma_{\text{norm}})_k$ and the true error level in $\delta_k$ by $(\sigma_{\text{true}})_k$ and define $\tau_k\equiv\frac{(\sigma_{\text{true}})_k}{(\sigma_{\text{norm}})_k}$. Since the noise is still assumed to be uncorrelated and zero-centered, the expectation value of the correlation is still zero, $\E[\rho_q]=0$. However, the variance of the correlations is proportional to the correlation of the squared error ratio, $\eta_q\equiv \frac{1}{N-q}\sum_{k=1}^{N-q}{\tau_k}^2{\tau_{k+q}}^2$,
  \begin{align}
    \Var[\rho_q] = \frac{\eta_q}{N-q}.
  \end{align}
  The expectation value of $\psisquaredred$ in this case is
  \begin{align}
    \E[\psisquaredred] = \frac{1}{N-1}\sum_{q=1}^{N-1}\eta_q.
  \end{align}
  This means that, in the absence of a non-random component, for imprecision in the noise level to have a negligible effect on results
  \begin{align}
    \left[\frac{1}{N-1}\sum_{q=1}^{N-1}(\eta_q-1)\right]^2 \ll \frac{14}{N}.
  \end{align}
  When the error is off by a constant factor, $\tau_k=\kappa$, this requirement yields $(\kappa^4-1)^2\ll\frac{14}{N}$. The corresponding requirement for $\chisquaredred$ is $(\kappa^2-1)^2\ll\frac{2}{N}$.

\section{$\psisquaredred$ for various residual features} \label{app:tables-of-values}

  Values of $\chisquaredred$ and $\psisquaredred$ for various curves calculated via Equations~\ref{eq:correlation-continuous}~and~\ref{eq:psisquaredred-continuous} are shown in Table~\ref{tab:nonrandom}. In general, a given non-random curve's $\chisquaredred$ is proportional to $A^2$ and independent of $N$, whereas its $\psisquaredred$ is proportional to $NA^4$.

  \begin{table}[h!!]
    \centering
    \begin{tabular}{|c|c|c|c|}
      \hline
      Feature & Form & $\chisquaredred$ & $\psisquaredred$ \\
      \hline
      Constant & $A$ & $A^2$ & $\frac{1}{2}\ NA^4$ \\
      Gaussian & $A\ \exp{\left[-\frac{1}{2}\left(\frac{\nu-\mu}{w}\right)^2\right]}$ & $\sqrt{\pi}\ A^2\left(\frac{w}{\Delta}\right)$ & $\sqrt{\frac{\pi^3}{2}}\ NA^4\left(\frac{w}{\Delta}\right)^3$ \\
      Harmonic & $A\ \sin{(k\nu+\phi)}$ where $k\Delta\gg 2\pi$ & $\frac{1}{2}\ A^2$ & $\frac{1}{16}\ NA^4$ \\
      Boxcar & $\begin{cases} A & (\nu-\mu)^2<(w/2)^2 \\ 0 & (\nu-\mu)^2>(w/2)^2 \end{cases}$ & $A^2\ \left(\frac{w}{\Delta}\right)$ & $\frac{1}{2}NA^4\  \eta\left(\frac{w}{\Delta}\right)$ \\
      Spike & $\begin{cases} A & \nu_k=\nu \\ 0 & \nu_k \neq \nu \end{cases}$ & $\frac{A^2}{N}$ & 0 \\
      \hline
    \end{tabular}
    \caption{Values of $\chisquaredred$ and $\psisquaredred$ for various non-random curves sampled at $N$ frequency channels between $\nu_{\text{min}}$ and $\nu_{\text{max}}$ (where $\Delta=\nu_{\text{max}}-\nu_{\text{min}}$). The values in the row corresponding to the Gaussian assume that the Gaussian is well-contained in the band (i.e. $\mu$ is many $w$'s away from either $\nu_{\text{min}}$ or $\nu_{\text{max}}$). The values in the table apply only when the data points are dense enough and broad enough that the given feature is well resolved. Here, $\eta(x) \equiv \left\{ x(3x-2)+4(1-x)^2\ \text{arccoth}\left(\frac{2}{x}-1\right) \right\}$ is a monotonically increasing function which satisfies $\eta(0)=0$, $\eta(1)=1$, and $\lim_{x\rightarrow 0^+}\frac{\eta(x)}{x^3}=\frac{2}{3}$. The `Spike' row can be viewed as the $\frac{w}{\Delta}\rightarrow\frac{1}{N}$ limit of the `Boxcar' row.} \label{tab:nonrandom}
  \end{table}

  \begin{table}[h!!]
    \centering
    \begin{tabular}{| r | c |}
      \hline
      \multicolumn{2}{|c|}{Assumption definitions} \\
      \hline
      A & $\E[\Delta_k]=0$ \\
      B & $\E[(\Delta_k)^2]=1$ \\
      C & $\E[(\Delta_k)^3]=0$ \\
      D & $\E[(\Delta_k)^4]=3$ \\
      \multirow{2}{*}{$\text{I}_n$} & $\E\left[\prod_k(\Delta_k)^{m_k}\right]=\prod_k\E[(\Delta_k)^{m_k}]$ \\ & 
      for $\{m_k\}\in\mathbb{N}^N$ s.t. $\sum_k m_k \le n$ \\
      \hline
    \end{tabular}
    \quad
    \begin{tabular}{| c | c | c | c | c | c | c |}
      \hline
      Result & Eq. ref. & A & B & C & D & $n$ \\
      \hline
      $\E[\rho_k]=0$ & \ref{eq:nonzero-correlation-moments} & \checkmark & & & & 2 \\
      $\Var[\rho_k]=\frac{1}{N-k}$ & \ref{eq:nonzero-correlation-moments} & \checkmark & \checkmark & & & 4 \\
      $\E[\psisquaredred]=1$ & \ref{eq:psisquaredred-nh-moments} & \checkmark & \checkmark & & & 4 \\
      $\Var[\psisquaredred]\simeq\frac{14}{N}$ & \ref{eq:psisquaredred-nh-moments} & \checkmark & \checkmark & \checkmark & \checkmark & 8 \\
      $\E[\chisquaredred] = 1$ & \ref{eq:chi-squared-moments} & \checkmark & \checkmark & & & 1 \\
      $\Var[\chisquaredred] = \frac{2}{N}$ & \ref{eq:chi-squared-moments} & \checkmark & \checkmark & & \checkmark & 4 \\
      \hline
    \end{tabular}
    \caption{Summary of the specific assumptions made to obtain various key results of this paper. The table on the left describes the different assumptions, while the table on the right shows which were assumed in each case. The quantity, $\bDelta$, is defined as $\bDelta=\bC^{-1/2}\bdelta$ in terms of the assumed covariance matrix $\bC$ and the residual vector $\bdelta$, as in Section~\ref{sec:definitions}.~The large sample limit, $\frac{p}{N}\rightarrow 0$, is used here. In the right table, an index $k$ implies $k\neq 0$. The $\mathbb{N}$ in the definition of assumption $\text{I}_n$ represents the set of non-negative integers.~All sums and products in the bottom row of the left table cover $1\le k\le N$. For the assumption $\text{I}_n$, $n$ is a measure of the independence of the components of $\bDelta$, with larger $n$ values representing stricter assumptions. $\text{I}_1$ represents no constraint while $\text{I}_\infty$ represents the assumption that all elements of $\bDelta$ are independent. If $\bdelta$ is Gaussian with mean 0 and covariance $\bC$, then A, B, C, D, and $\text{I}_\infty$ are true.}
    \label{tab:gaussianity-assumptions}
  \end{table}
  
\section{Detailed assumptions} \label{app:gaussianity-assumptions}

  For the sake of simplicity, throughout this paper, our explicit assumption has been that the residual $\bdelta$ is Gaussian.~However, this strong assumption is not necessary in general. Table~\ref{tab:gaussianity-assumptions} enumerates the assumptions made in order to reach various key results of the NH distributions of $\chisquaredred$ and $\psisquaredred$.~As shown in this table, the formula for the variance of any quantity usually requires a larger set of assumptions than that for the mean of the same quantity. If, in a given situation, an assumption in Table~\ref{tab:gaussianity-assumptions} is not true, the corresponding result should be recalculated accordingly.

\acknowledgments

We thank the referee for their helpful remarks. This work was directly supported by the NASA Solar System Exploration Research Virtual Institute cooperative agreement 80ARC017M0006 and by NASA ATP grant NNX15AK80G to JB. DR is supported by a NASA Postdoctoral Program Senior Fellowship at the NASA Ames Research Center, administered by the Universities Space Research Association under contract with NASA.

\bibliographystyle{JHEP}
\bibliography{manuscript}

\end{document}